%% file: ANA-HION-2023-03-PAPER.tex
\documentclass[PAPER, cernpreprint=true, atlasdraft=false, UKenglish, texmf, orcidlogo]{atlasdoc}
\usepackage{atlaspackage}

\usepackage{atlasbiblatex}

\usepackage{ANA-HION-2023-03-defs}

\usepackage{atlasphysics}
\graphicspath{{logos/}{figures/}}

\AtlasTitle{Measurement of isolated photon plus two-jet correlations in Pb+Pb and $pp$ collisions at 5.02~TeV with ATLAS}

\AtlasAbstract{%
This paper presents a measurement of
photon plus two-jet events in $pp$ and Pb+Pb collisions, i.e. events in which the transverse momentum of a single photon is balanced by two distinct jets.
The measurement was performed using $pp$ data taken in 2017 with an integrated luminosity 260~pb$^{-1}$, and Pb+Pb data taken in 2018 with an integrated luminosity 1.72~nb$^{-1}$, both at $\sqrt{s_\mathrm{NN}}$ =5.02~TeV, as recorded by the ATLAS detector.
Events with photons in the transverse momentum range 90--180~GeV and at least two anti-$k_t$ $R = 0.2$ jets with a $p_\mathrm{T}$ > 30~GeV are selected, and three observables are measured: ${\mathrm{x}}_{\mathrm{JJ}\gamma}$, $\mathrm{A}_{\mathrm{JJ}\gamma}$, and $\Delta R_{\mathrm{JJ}}$. These observables characterise the overall energy loss of the multiparton system from medium interactions (${\mathrm{x}}_{\mathrm{JJ}\gamma}$), the relative energy loss between the two colour-charge carriers ($\mathrm{A}_{\mathrm{JJ}\gamma}$), and medium-induced modifications to their opening angle ($\Delta R_{\mathrm{JJ}}$). The observables are corrected for uncorrelated combinatoric background contributions using a novel multijet mixing technique, for photon purity, and for detector resolution effects via iterative unfolding. Final results are presented per photon, and the ratio ($I_\mathrm{AA}$) is taken between measurements in Pb+Pb and $pp$ collisions, for Pb+Pb centrality intervals of 30--80\%, 10--30\%, and 0--10\%. Significant suppression of per photon two-jet yields in all three observables, $I_\mathrm{AA} < 1$, is observed as a result of parton-medium interactions.   The experimental measurements are compared to three different jet quenching models:
\textsc{Jewel}, \textsc{Jetscape}, and \textsc{Lbt}.
}
\AtlasRefCode{HION-2023-03}

\PreprintIdNumber{CERN-EP-2026-121}

\AtlasJournal{Phys.\ Lett.\ B.}
\hypersetup{pdftitle={ATLAS document},pdfauthor={The ATLAS Collaboration}}

\newcommand{\AtlasCoordFootnote}{%
ATLAS uses a right-handed coordinate system with its origin at the nominal interaction point (IP)
in the centre of the detector and the \(z\)-axis along the beam pipe.
The \(x\)-axis points from the IP to the centre of the LHC ring,
and the \(y\)-axis points upwards.
Cylindrical coordinates \((r,\phi)\) are used in the transverse plane,
\(\phi\) being the azimuthal angle around the \(z\)-axis.
The pseudorapidity is defined in terms of the polar angle \(\theta\) as \(\eta = -\ln \tan(\theta/2)\).
Angular distance is measured in units of \(\Delta R \equiv \sqrt{(\Delta\eta)^{2} + (\Delta\phi)^{2}}\).}

\begin{document}

\maketitle

\tableofcontents

\section{Introduction}

Collisions of ultra-relativistic heavy ions at the Large Hadron Collider (LHC) and the Relativistic Heavy Ion Collider (RHIC) produce a state of matter called the quark--gluon plasma (QGP) where quarks and gluons exist outside of the colour-neutral configurations that define typical cold matter. Studies of this hot-and-dense  state of deconfined partons have revealed many interesting characteristics: collective motion well described by relativistic hydrodynamic flow, sequential suppression of bound quarkonia states, and the suppression of observed final-state products such as jets from hard-scattered partons~\cite{Busza:2018rrf}. This last phenomenon was first observed at RHIC with the study of large transverse momentum (\pt) hadrons~\cite{PHENIX:2001hpc} and later at the LHC with fully reconstructed jets~\cite{HION-2013-06}. It is interpreted as the result of the colour-opaque medium interacting with and attenuating the energy of these energetic quarks and gluons produced in the initial collision before the formation of the medium, a process referred to as jet quenching. For a recent review of jet quenching studies in the QGP at both colliders, see Ref.~\cite{Cunqueiro:2021wls}.

Measurements of jet quenching are a key part of understanding the structure and nature of the QGP\@. By comparing the kinematics of jets produced in Pb+Pb collisions to the well-measured characteristics of jets produced in \ppp collisions, the induced modifications from the QGP interacting with the developing parton shower can be studied. While many studies of modified jet production have been performed~\cite{HION-2017-10,HION-2019-02,HION-2020-09,CMS-HIN-14-010,CMS-HIN-16-006,CMS-HIN-18-014,ALICE:2015mjv}, there are unresolved questions on the exact nature of parton--medium interactions. These include questions on how much energy is lost to interactions with medium partons, how the energy loss depends on parton flavour, and whether there is a minimum separation between colour charges before the medium can resolve them as separate and independent~\cite{Casalderrey-Solana:2012evi,Casalderrey-Solana:2019ubu,HION-2021-09}.

The measurements presented here address these questions by studying events containing an isolated photon with its transverse momentum balanced by at least two distinct jets, $\gamma$ + 2 jets + $X$. As the photon has no colour charge, it sets the scale of the initial hard scattering that produced it and the balancing two-jet system, without any biases on this selection from quenching effects. Photon plus two-jet measurements were previously reported by ATLAS in \ppp collisions at 13~\TeV~\cite{STDM-2017-32}.
Measurements of photon-tagged inclusive jets, i.e. $\gamma$ + 1 jet + $X$, in \ppp and Pb+Pb collisions, were performed previously by ATLAS~\cite{HION-2018-06,HION-2022-14} and CMS~\cite{CMS-HIN-16-002}.   This analysis presents the first photon plus two-jet measurement in Pb+Pb collisions.

The multijet system can produce one quark and one gluon jet opposite the photon, such that measurements of the jet--jet energy asymmetry probe if different colour charges have different interactions with the medium.
The measurement is also sensitive to fluctuations in the energy loss as the two jets traverse similar regions of QGP, in contrast to the case for back-to-back dijets.   Finally, studies as a function of the angle between the jet pair can be used to probe the existence of the colour coherence regime.
Although some estimates predict such coherence effects only at very small angular separations, and there are existing measurements probing that phase space~\cite{HION-2021-09,ALargeIonColliderExperiment:2021mqf,CMS-HIN-23-001}, the new data presented in this paper can provide constraints on possible effects at larger angular separation.

This measurement considers three new observables defined by the photon and all jet pairs opposite the photon in azimuth. Within a given photon+jet pair, ``1'' or ``leading'' denotes the jet of higher \pt, and ``2'' or ``subleading'' denotes the jet of lower \pt~\footnote{\AtlasCoordFootnote}.
The observables are the ratio of the magnitude of the two-jet vector \pt to that of the photon (\xjjg), the difference between the two jet \pt values divided by that of the photon (\ajjg), and the angle between the jet pair (\drjj), in events containing a photon and at least two jets. More explicitly, these are defined as:
\begin{equation}
\label{eqn:xjjg}
\xjjg = (\vec{p}_{1} + \vec{p}_{2})_{\mathrm{T}}/\ptg,
\end{equation}
\begin{equation}
\ajjg = (p_{\mathrm{T},1} - p_{\mathrm{T},2})/\ptg,
\end{equation}
\begin{equation}
\label{eqn:drjj}
\drjj = \sqrt{\Delta\phi_{1,2}^2 + \Delta\eta_{1,2}^2}.
\end{equation}
In events containing more than two candidate jets, all possible combinations of the photon with two jets are considered separately.  Thus, a given jet can take on the leading or subleading role in the different $\gamma + 2$ jets $+X$ combinations. The average number of jets ranges from 2.2 -- 2.3 depending on the Pb+Pb centrality interval.  The distributions are normalised to the number of photons.

The measurements presented are performed using \ppp data taken in 2017 corresponding to an integrated luminosity of 260~\pbi and Pb+Pb data taken in 2018 corresponding to an integrated luminosity of 1.72~\nbi, both at $\sqrt{s_\mathrm{_{NN}}}$ = $5.02$~\TeV.
Events that contain an isolated photon with transverse momentum of $90 < \ptg <180$~\GeV and at least two
jets in the azimuthally opposite hemisphere
($\dphijg = \left|\phi^\mathrm{jet} - \phi^\gamma\right| > \pi/2$)
with $\pt > 30$~\GeV are studied.
A small jet radius parameter $\textit{R}=0.2$ is chosen to enable the reconstruction of jet pairs as close as possible in $\Delta \textit{R}$, and also to reduce jet background contributions in Pb+Pb collisions.
In addition to jets specifically associated with the initial hard process that produced the photon, the Pb+Pb events contain combinatorial background, meaning jets arising from uncorrelated hard-scatterings or underlying event (UE) fluctuations. A novel multijet mixed-event procedure, developed specifically for this analysis, is employed to correct for these contributions in the data. The three observables \xjjg, \ajjg, and \drjj are unfolded to the particle-level and reported in \ppp and Pb+Pb collisions. The ratios of the Pb+Pb distributions to those in \ppp collisions, referred to as \IAA, are used to characterise parton energy loss in the QGP and are compared with theoretical models.

\section{ATLAS detector}
The ATLAS experiment~\cite{PERF-2007-01} at the LHC is a multipurpose particle detector
with a forward--backward symmetric cylindrical geometry and a near \(4\pi\) coverage in
solid angle.
It consists of an inner tracking detector surrounded by a thin superconducting solenoid
providing a \qty{2}{\tesla} axial magnetic field, electromagnetic and hadron calorimeters, and a muon spectrometer.
The inner tracking detector covers the pseudorapidity range \(|\eta| < 2.5\).
It consists of silicon pixel, silicon microstrip, and transition radiation tracking detectors.
Lead/liquid-argon (LAr) sampling calorimeters provide electromagnetic (EM) energy measurements
with high granularity.
A steel/scintillator-tile hadron calorimeter covers the central pseudorapidity range (\(|\eta| < 1.7\)).
The endcap and forward regions are instrumented with LAr calorimeters
for both the EM and hadronic energy measurements up to \(|\eta| = 4.9\).
The muon spectrometer surrounds the calorimeters and is based on
three large superconducting air-core toroidal magnets with eight coils each.
The field integral of the toroids ranges between \num{2.0} and \qty{6.0}{\tesla\metre}
across most of the detector.
The muon spectrometer includes a system of precision tracking chambers and fast detectors for triggering.
Two zero-degree calorimeters (ZDCs) were situated at $\left|\eta\right| > 8.3$ during Pb+Pb data-taking and is mostly sensitive to spectator neutrons from the fragmenting nuclei in Pb+Pb collisions.

A two-level trigger system is used to select events.
The first-level trigger is implemented in hardware and uses a subset of the detector information
to accept events at a rate below \qty{100}{\kHz}.
This is followed by a software-based trigger that
reduces the accepted event rate to \qty{1}{\kHz} on average
depending on the data-taking conditions.
An extensive software suite~\cite{ATL-SOFT-PUB-2021-001} is used in data simulation, in the reconstruction
and analysis of real and simulated data, in detector operations, and in the trigger and data acquisition
systems of the experiment.

\section{Data selection, reconstruction, and simulation}
\label{sec:dataMC}

The data samples, physics object reconstruction, and simulation samples used in this measurement are identical to those used in a previous measurement of photon-tagged jet production~\cite{HION-2022-14}, and are briefly summarised here.

Events in data are selected using triggers requiring a reconstructed photon with $p_\mathrm{T}$ above 35~\GeV (20~\GeV) in \ppp (Pb+Pb) collisions~\cite{TRIG-2016-01,TRIG-2018-05}. These triggers sample the full luminosity delivered in both 2017 \pp and 2018 Pb+Pb data and are fully efficient for the photon selection used in this analysis.
Events containing multiple Pb+Pb interactions within the same or nearby bunch crossings are rejected using an offline selection based on the correlation between signals measured in the forward calorimeters and the ZDCs.
Photons are reconstructed following the method used previously in Pb+Pb measurements~\cite{HION-2018-06,HION-2017-08}. This procedure is the same as that used in \pp\ collisions~\cite{EGAM-2018-01} after an event-by-event estimate and subtraction of the UE contribution to the energy deposited in each calorimeter cell~\cite{HION-2011-02}. Photon candidates must satisfy shower-shape requirements~\cite{PERF-2017-02}
designed to reject photons from neutral meson decays and other hadrons.
Photon candidates must also be isolated by requiring an upper limit of 3~\GeV on the sum of the transverse energy (after subtracting the contributions from the photon itself and the UE) in calorimeter cells within $\Delta{R} < 0.3$ from the photon.

Jets are reconstructed following the procedure previously used in Pb+Pb measurements~\cite{HION-2011-02,HION-2017-10}. The \antikt algorithm~\cite{Cacciari:2008gp,Cacciari:2011ma} with radius parameter $\textit{R}=0.2$ is applied to cells in all calorimeter layers combined into logical towers of size $\Delta\eta\times\Delta\phi$ = $0.1\times\pi/32$.
The contribution to the energy deposited in towers by the UE is estimated on an event-by-event basis, and the tower kinematics are iteratively updated to subtract the UE contribution that is then re-estimated. The resulting energy response of the jets is corrected using simulation to account for the response of the calorimeter to jets~\cite{JETM-2019-01}.  Further corrections are applied using {\em in situ} studies of jets recoiling against photons, $Z$ bosons, and jets in other regions of the calorimeter in \pp\ collisions~\cite{HION-2018-06} for the absolute response in data.

The Pb+Pb event centrality is defined by the total sum of the transverse energy in the forward calorimeters, $\Sigma{E}_\mathrm{T}^\mathrm{FCal}$. Events in different ranges of $\Sigma{E}_\mathrm{T}^\mathrm{FCal}$ are associated with an underlying geometric configuration according to a Monte Carlo (MC) Glauber simulation~\cite{Miller:2007ri} using the same event-selection criteria as in previous ATLAS analyses~\cite{HION-2016-07}.
Three centrality intervals are utilised corresponding to the following fractions of the $\Sigma{E}_\mathrm{T}^\mathrm{FCal}$ distribution in minimum-bias events: 0--10\% (``central'' events, with a large nuclear overlap), 10--30\%, and 30--80\% (``peripheral'' events).

Monte Carlo simulations of photon+jet events are used to evaluate the performance of the photon and jet reconstruction and to correct the measured distributions for detector effects. The distributions are thus corrected to the ``particle level.''

Particle-level photons and jets are reconstructed before detector simulation. Photons are required to be prompt and to satisfy $|\eta^\gamma| < 2.37$
(excluding the barrel-to-endcap transition region of $1.37 < |\eta^\gamma| < 1.52$)
and $90 < \ptg < 180$~\GeV. Truth isolation for photons is defined as the sum of transverse energy within a cone of $\Delta R < 0.4$ around the photon being less than 5~\GeV. All final-state particles, excluding neutrinos, muons, and the photon itself, are used to calculate this quantity at the truth level.

Particle-level jets are clustered from all detector-stable particles with a lifetime in the laboratory frame satisfying $c\tau > 10$~mm, excluding those expected to deposit little or no energy in the calorimeters (i.e., muons and neutrinos). Particle-level jets are required to satisfy $|\eta_{\textrm{jet}}| < 2.8$ and $\pt > 30$~\GeV.

For \pp data, the main MC sample consists of \pythia-8~\cite{Sjostrand:2007gs} events. These include contributions from both direct and fragmentation photons and are produced using the A14 tune~\cite{ATL-PHYS-PUB-2014-021} and the NNPDF~2.3 LO parton distribution function (PDF) set~\cite{ATL-PHYS-PUB-2014-021}.
These events are simulated~\cite{ATL-SOFT-PUB-2021-001} within ATLAS using \GEANT~\cite{GEANT4:2002zbu} and are digitized and reconstructed in a manner identical to that in data. To simulate photon+jet events in Pb+Pb collisions, these \pythia-8 events are overlaid at the detector-hit level with a sample of Pb+Pb data events, and the combination is reconstructed as a single event.
These ``Pb+Pb data overlay'' events are reweighted to match the observed $\Sigma{E}_\mathrm{T}^\mathrm{FCal}$ distribution for photon+jet events in Pb+Pb data.
\section{Analysis}
\label{sec:analysis}

The signal definition for the measurement presented here requires events having an isolated photon in the range of $90 < \ptg < 180$~\GeV and $|\eta^\gamma| < 2.37$ (excluding the barrel-to-endcap transition region of $1.37 < |\eta^\gamma| < 1.52$), and at least two $\textit{R}=0.2$ jets with $ \pt > 30$~\GeV and $|\eta^\mathrm{jet}| < 2.8$ that are in the azimuthal hemisphere opposite to the photon, $\dphijg = \left|\phi^\mathrm{jet} - \phi^\gamma\right| > \pi/2$. For each event satisfying this initial selection, the photon together with every possible combination of two jets is considered as a set.
For each set, three additional criteria are required: (1) both jets must be separated from the photon by $\Delta R>0.5$, (2) the two jet axes must be separated by $\drjj > 0.4$, and (3) the azimuthal separation between the two-jet system and the photon must have $\dphijjg > 7\pi/8$.
Since at tree level a prompt photon plus two partons are expected to have approximately balanced momenta in the transverse plane, this last requirement reduces the rate of the combinatorial contributions while minimally impacting the number of signal events.
With these requirements, approximately 144,000 and 34,000 events are selected in \ppp and Pb+Pb data sets, respectively.
Then initial distributions are constructed for the three new observables \xjjg, \ajjg, and \drjj as defined in Eqs.~\ref{eqn:xjjg}-\ref{eqn:drjj}.

In the candidate Pb+Pb events, a subset of the selected jets do not arise from the same hard scattering as the photon, but rather from an unrelated nucleon–nucleon scattering, or from jets that are reconstructed from localised fluctuations of the UE\@. These background jets are produced predominantly at low \pt and isotropically relative to the photon direction. Such background jets can influence the observables by forming a signal-jet + background-jet pair, or even background-jet + background-jet pair.
The influence of these jets on the observables is removed statistically using a \textit{mixed-event subtraction} procedure.
A simpler form of this procedure was previously used for photon + inclusive jet analyses~\cite{HION-2018-06,HION-2022-14}.
Due to the complexity of multiple jets contributing to the observables, this technique was developed further.

For each signal event (consisting of a photon and at least two jets), two minimum-bias
Pb+Pb data events are chosen that have similar centrality and second-order event plane angle~\cite{HION-2019-11} as the signal event, and thus have on average the same distribution of background jets. The distributions are then constructed using three different combinations of the photon and jets in the signal event and the jets in the minimum-bias events, as follows:

\begin{enumerate}
\item First, the contribution from background-jet + background-jet pairs that occur in data must be accounted for. The observable of choice is constructed using the photon from the signal event in combination with all possible jet pairs in the first minimum-bias event.
\item The second step accounts for signal-jet and background-jet combinations. For each individual jet satisfying selections in the signal event, the observable is constructed using the photon and that one jet in combination with all the jets in the first minimum-bias event.
However, since the jet from the signal event may itself be a background jet, this estimated background will generally include background-jet (from the signal event) and background-jet (from the first minimum bias event) pairs. Since these background--background pairs are from different minimum-bias events, this contribution is not present in data. %
\item The third step accounts for the potential over-subtraction described in the second step. Here, the observable of choice is constructed from the photon in the signal event, each jet in the first minimum-bias event satisfying selections, and each jet in the second minimum-bias event satisfying selections. If the first minimum-bias event contains two jets, and the second minimum-bias event contains three jets, this results in six total pairs, assuming no pair is excluded by jet--jet selections like the $\Delta \textit{R} > 0.4$ selection. This contribution accounts for the background-jet plus background-jet pairs from different events that are introduced by the second step.
\end{enumerate}

While this procedure is described for a single minimum-bias-event pair per signal event, the average of ten minimum-bias-event pairs per signal event was employed in the analysis to improve statistical precision.  For each observable, the per-photon yields from steps one and two are subtracted from the initial yields, and the yields from step three are added (to correct for the background over-subtraction), to produce a set of combinatorial-background-subtracted distributions.
The procedure is tested using the \pythia-8 and Pb+Pb data overlay events (described in Sec.~\ref{sec:dataMC}), where reconstructed jets matched to a \pythia-8 particle-level jet are taken to be the signal jets, and the other jets in the event, i.e. those in the Pb+Pb data event, are the background jets.
Complications such as the potential merging of jets at reconstruction level results in some residual non-closure of the background procedure, which is taken as a systematic uncertainty in the measurement (discussed in Sec.~\ref{sec:syst}). The observed background correction in Pb+Pb data is of comparable magnitude to the signal in the central events and is greater at large \ajjg, large \drjj, and the tails of the \xjjg distribution. In these regions the non-closure and corresponding systematic uncertainty can be as large as 20\%.

An additional background arises from photons produced in decays of neutral meson and other hadrons. The purity of the selected photons is estimated following the data-driven, double-sideband method used in ATLAS photon measurements~\cite{STDM-2014-09,STDM-2016-08,STDM-2012-16,STDM-2015-11}.   The obtained purity ranges from $80$--$95$\% depending on the collision system and \ptg. The contribution to the jet--jet combinations from this background is estimated in data and statistically subtracted using the same methods applied in previous measurements~\cite{HION-2018-06,HION-2017-08,HION-2022-14}.   The method is briefly summarised here. An inverted set of photon selection requirements designed to enhance the contribution from background is used to select events, and the analysis is repeated, including the combinatorial background subtraction. The resulting jet--jet yields in these background events are scaled according to the estimated photon purity and are statistically subtracted from the yields obtained with the nominal photon selection.

An unfolding procedure is applied to the data to correct for the kinematic-bin migration introduced by detector resolution and UE effects and for the finite reconstruction and selection efficiency. This correction is performed using the iterative Bayesian method~\cite{DAgostini:1994fjx} in the \roounfold~\cite{Adye:2011gm} software package. The simulation samples described in Sec.~\ref{sec:dataMC} are used to produce response matrices in \ppp and Pb+Pb collisions.
The response matrices are observed to be predominantly diagonal and well-behaved, demonstrating limited migration between neighbouring bins.
The simulation samples are also used to test the unfolding procedure by using half of the simulation to construct response matrices while treating the other half as data to be unfolded and compared with particle level kinematics. For each of the three observables, the unfolding is performed in three dimensions: the given observable (for example, \xjjg), \ptg, and $p_{\mathrm{T},2}$. The latter two variables are included since the response along the observable axis is found to be particularly dependent on their values. The kinematic distributions are constructed in a wider range than the region in which the final results are reported, to allow for the unfolding to properly account for the migration into and out of the measurement region.
Before the unfolding procedure is applied, the simulated events are first reweighted to better describe the distributions as observed in data.
The weights are applied at the particle-level, as a function of \ptg and $p_{\mathrm{T},2}$, and the specific observable, on a photon-jet pair basis.
The number of iterations is chosen by minimizing the quadrature sum of the statistical uncertainties with the change in the distributions from the previous iteration, and is typically two or three depending on the observable and event class.

\section{Systematic Uncertainties}
\label{sec:syst}

The sources of systematic uncertainty in this measurement are those associated with the measurement of the photons and the jets,  the background subtraction procedure, and the unfolding. The uncertainties are typically evaluated by repeating the full analysis chain with a given systematic variation, which may result in, e.g., a different response matrix for the unfolding or a different reconstructed-level distribution.

The photon- and jet-related sources of uncertainty are the same as those used in a previous measurement of photon-tagged jet production in Pb+Pb and \ppp collisions~\cite{HION-2022-14} and are only summarised here. The photon-related uncertainties include those on the photon energy scale and resolution~\cite{PERF-2017-03}, the application of identification and isolation requirements~\cite{PERF-2017-02}, and the evaluation of the photon purity~\cite{HION-2018-06}. The jet-related uncertainties include the baseline uncertainties in the energy scale and resolution for $R=0.4$ jets used for measurements in 13~\TeV{} \ppp collisions~\cite{JETM-2019-01}, with additional sources~\cite{HION-2019-02} accounting for the modified jet reconstruction procedure used in a heavy-ion environment, the smaller jet parameter $R=0.2$, the data-taking conditions in $\sqrt{s_\mathrm{_{NN}}}$ = 5.02~\TeV running, the quark/gluon composition of photon plus two-jet events at this collision energy, and the potential impact of jet quenching on the calorimeter response to jets.

The uncertainty associated with the mixed-event background subtraction procedure, present only for Pb+Pb collisions, is evaluated using input from the closure test of the procedure in the Pb+Pb data overlay simulation described in Sec.~\ref{sec:analysis}. For each of the three observables, the differences between the background-subtracted distribution and the reconstructed-level signal distribution in simulation are applied to the data bin-by-bin. The unfolding is then performed on these modified input distributions.

The uncertainty associated with the unfolding procedure has three components. The first is the sensitivity to the initial distributions used in the iterative unfolding, which is evaluated by using the simulation samples without reweighting to the data. Second, the reconstructed jet \pt threshold is raised from 30~\GeV to 35~\GeV, but still unfolded to the original kinematic range. Such a variation was used in Ref.~\cite{HION-2019-02} as a check against residual background jet contamination at low jet \pt. Finally, the impact of finite statistical uncertainty in the simulation sample is evaluated with a pseudo-data re-sampling technique implemented in the \roounfold package~\cite{Adye:2011gm}. This last component is much smaller than the statistical uncertainties in the data.

For the measurements in Pb+Pb data, the total uncertainties for each observable are typically 20--40\%, but reach as high as 100\% in the low-yield tails. The photon isolation and purity, the combinatorial background subtraction, and the unfolding uncertainties have similar values and are co-dominant. For the \ppp measurement, the total uncertainties are typically 5--10\%, with the jet energy scale and the unfolding uncertainties (which include an observed unfolding non-closure from a simulation study) being dominant. Variations for sources common to the Pb+Pb and \ppp data are considered together when evaluating \IAA, the ratio of distributions in these systems. This results in a modest reduction of the uncertainties in \IAA compared with the relative uncertainties in the \ppp or Pb+Pb distributions alone.

\begin{figure}[t!]
\begin{center}
\subfigure[]{\includegraphics[width=0.32\textwidth]{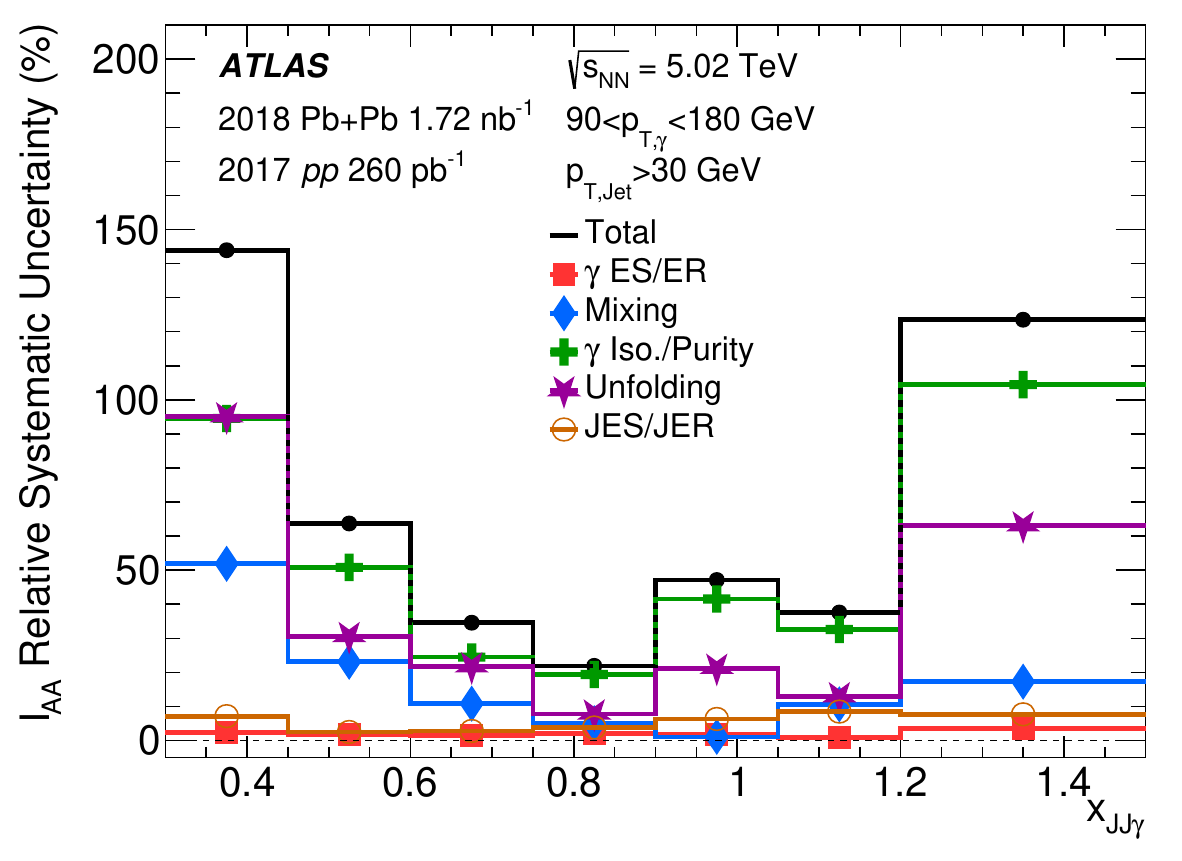}}
\subfigure[]{\includegraphics[width=0.32\textwidth]{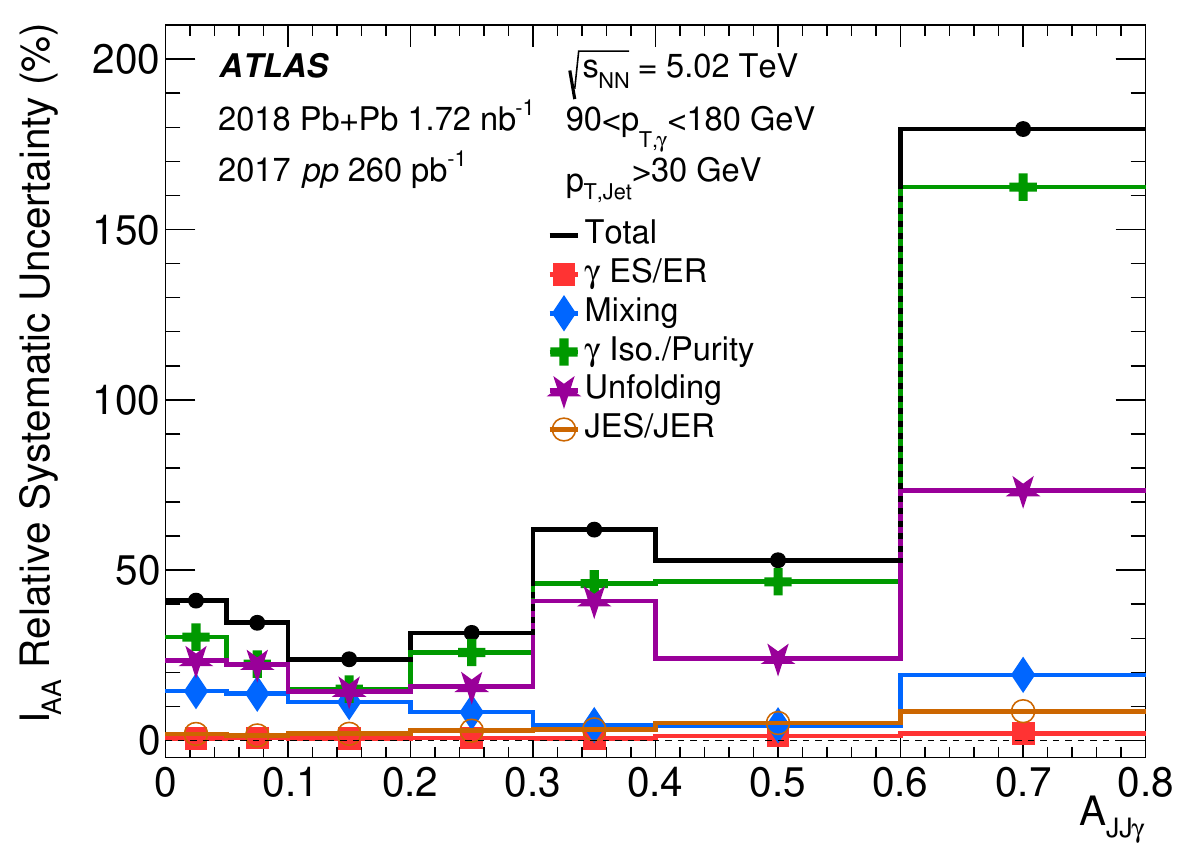}}
\subfigure[]{\includegraphics[width=0.32\textwidth]{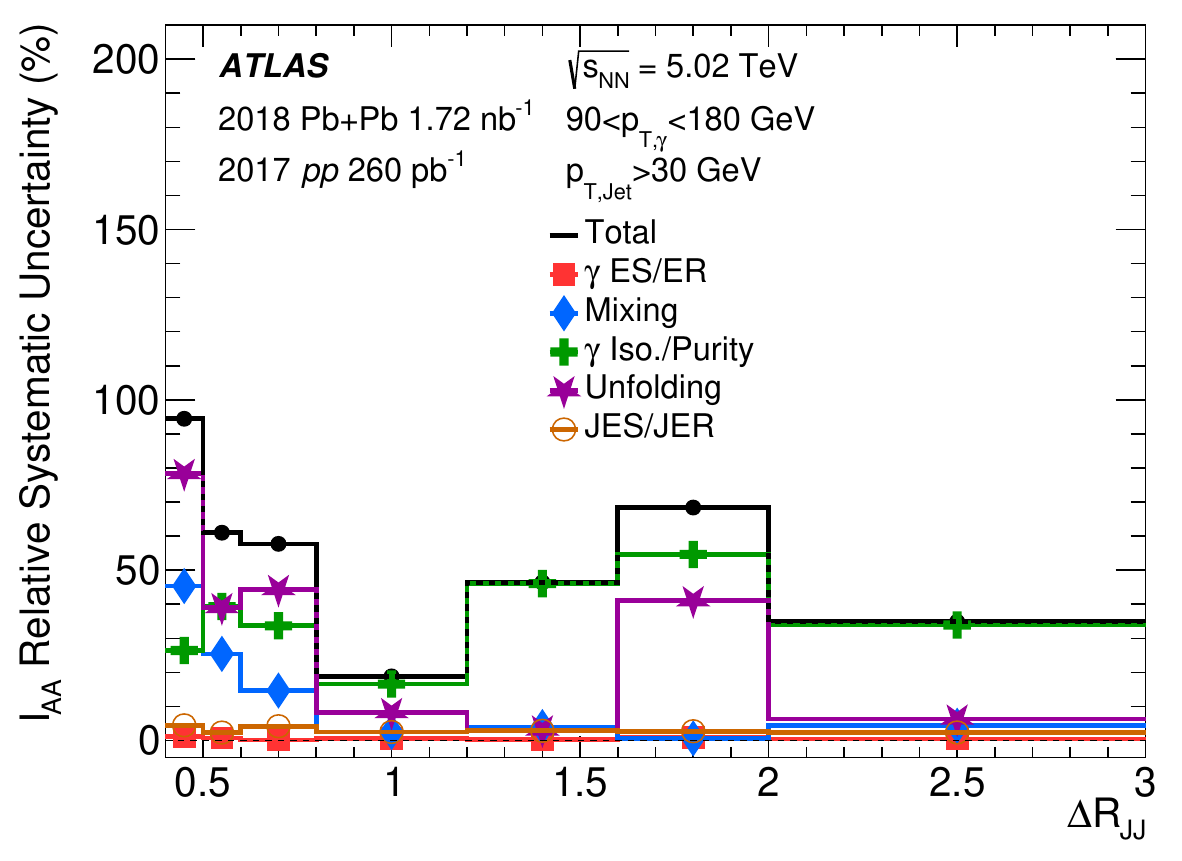}}
\caption{
Relative systematic uncertainties shown as percentages on \IAA, the ratio of the (a) \xjjg, (b) \ajjg, and (c) \drjj distribution between  central Pb+Pb and \ppp events.
The total systematic uncertainty is represented by the black histogram with filled black circles. Correlations in systematic uncertainties between the Pb+Pb and \ppp yields are accounted for in evaluating the ratio.}
\label{fig:systCent0to10OverPP}
\end{center}
\end{figure}

Figure~\ref{fig:systCent0to10OverPP} shows the contributions of various sources to the systematic uncertainty in \IAA, the ratio of central Pb+Pb to \pp{} data, in each of the three observables. Overall, the dominant uncertainties originate from
photon isolation and purity ($\gamma$ Iso/Purity),
unfolding (Unfolding),
and mixing non-closure (Mixing),
with additional contributions coming from jet energy scale/resolution (JES/JER) and the photon energy scale/resolution ($\gamma$ ES/ER).
In the tails of the distributions, the total relative systematic uncertainty can exceed 100\%.

\section{Results}

\begin{figure}[hbtp!]
\begin{center}
\subfigure[]{\includegraphics[width=0.32\textwidth]{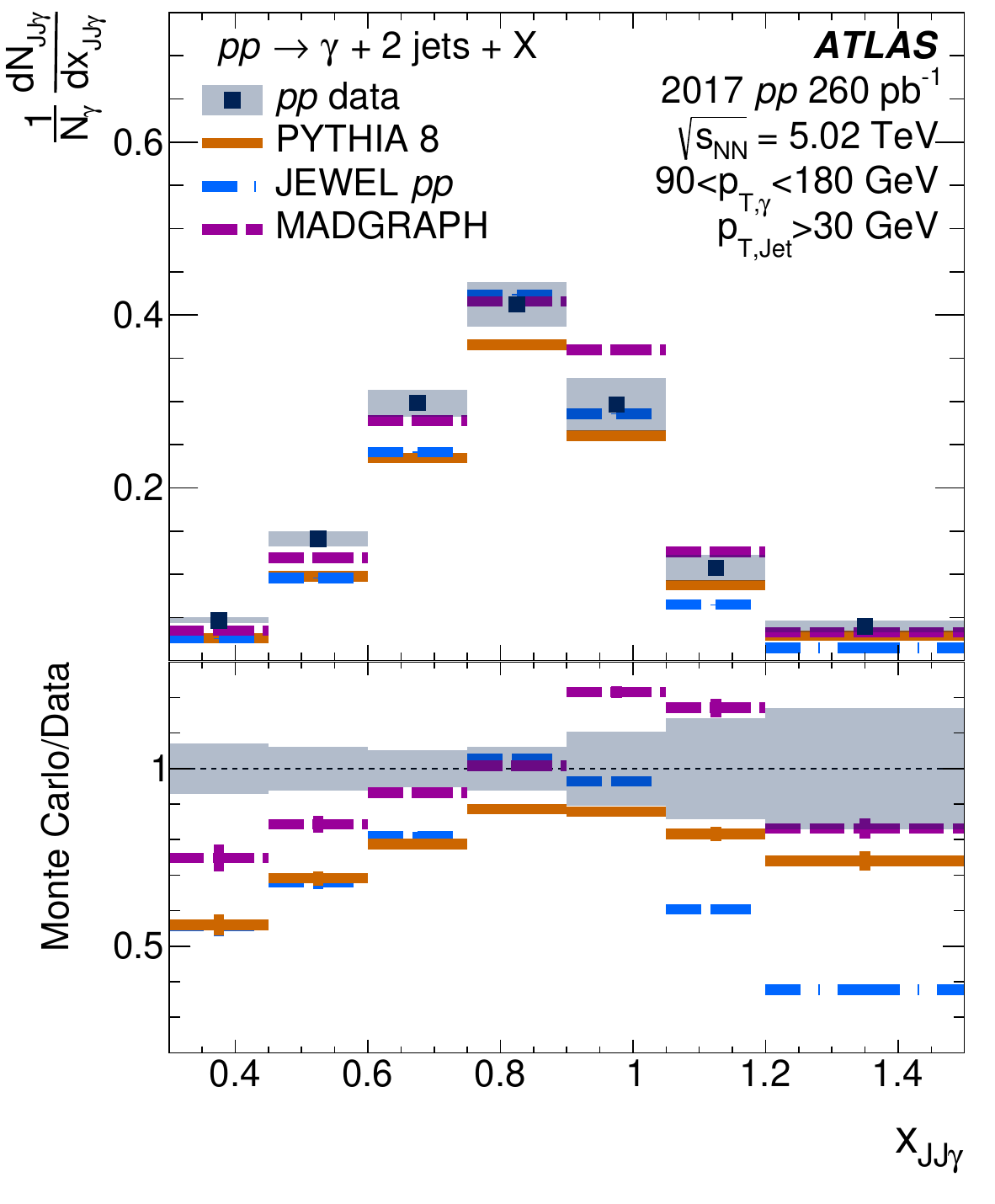} }
\subfigure[]{\includegraphics[width=0.32\textwidth]{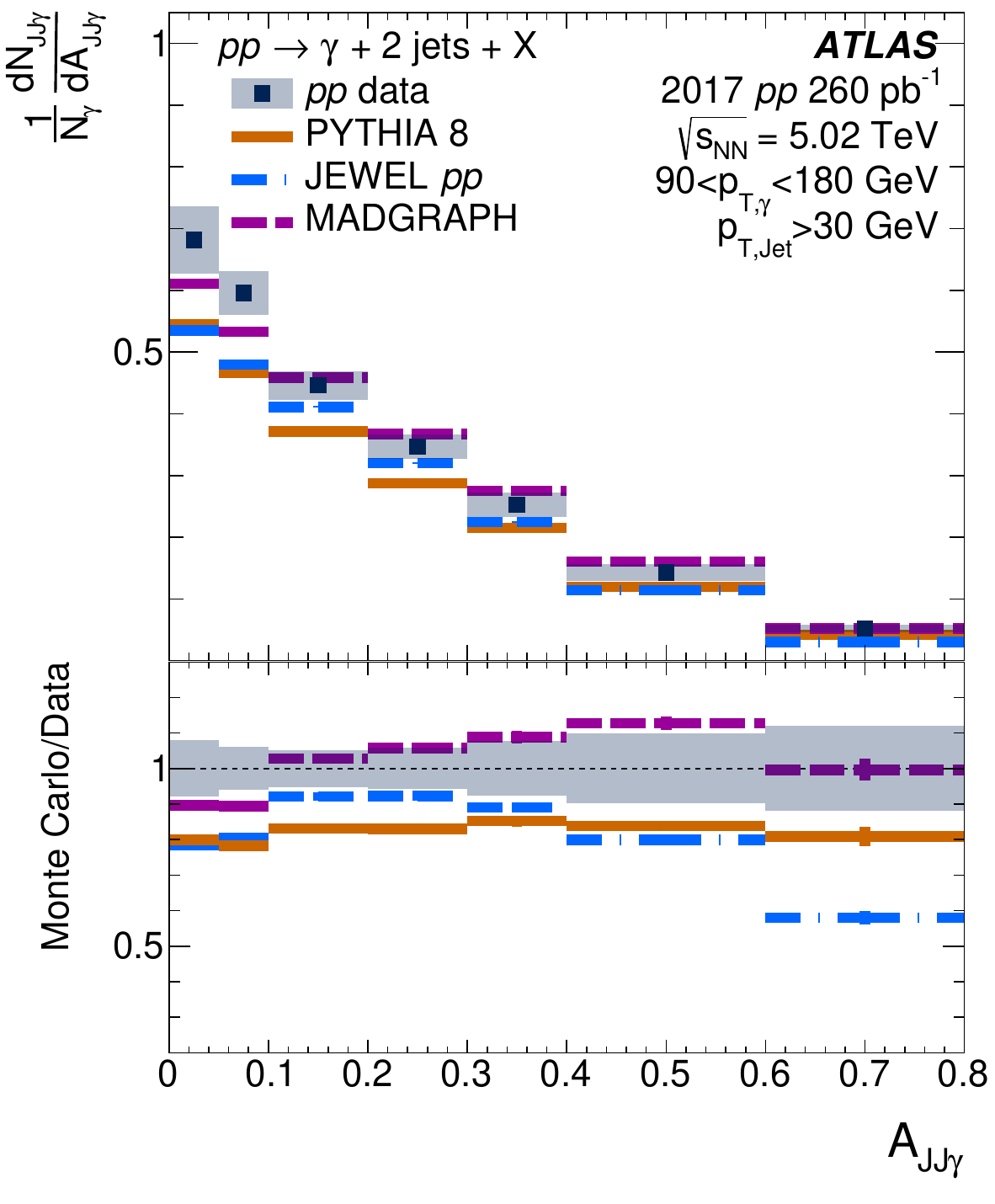}}
\subfigure[]{\includegraphics[width=0.32\textwidth]{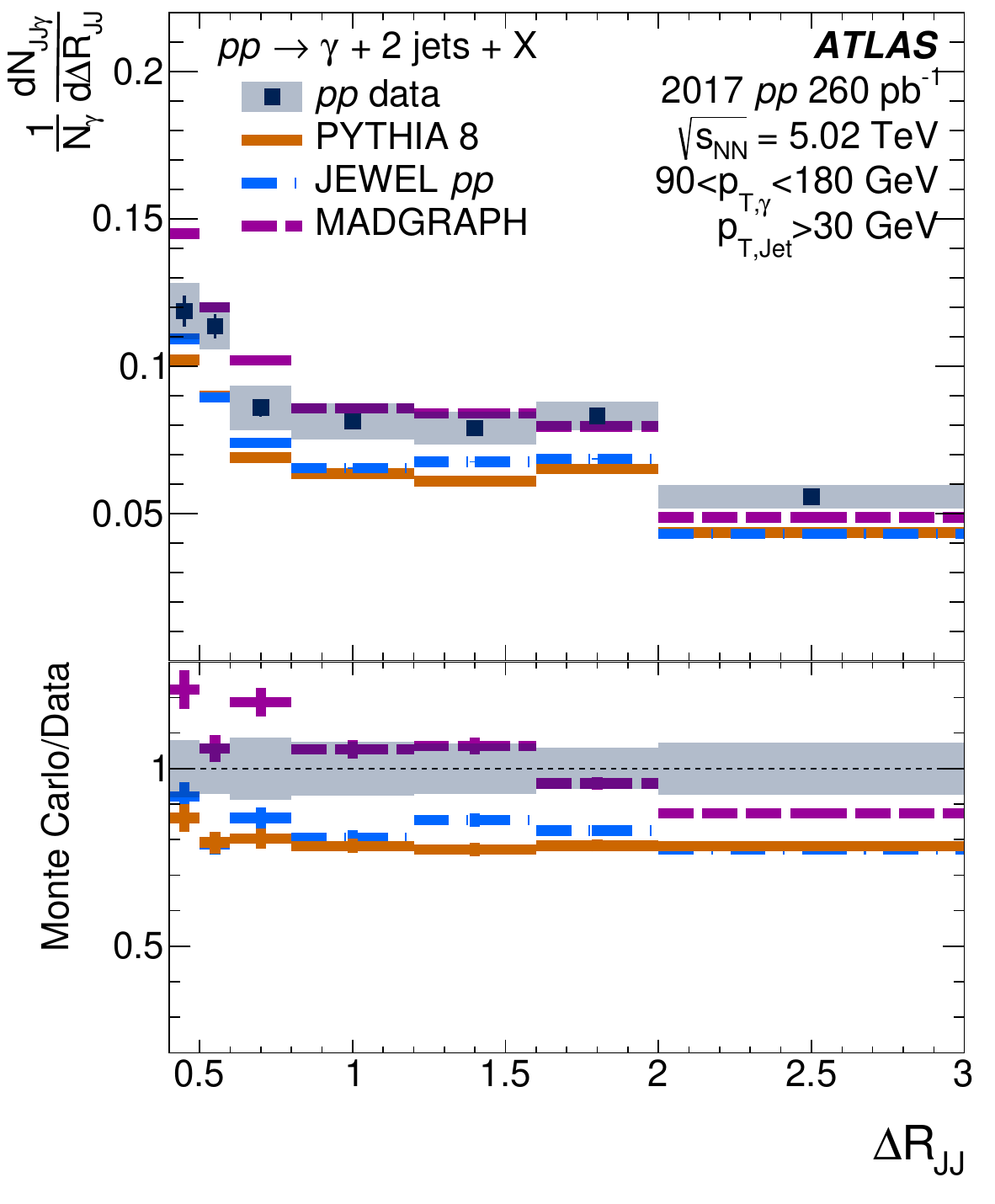}}
\caption{
Unfolded data distributions in \ppp collisions (a) \xjjg, (b) \ajjg, and (c) \drjj. MC generator predictions by the \pythia-8, \jewel pp, and \MADGRAPH generators are shown.
The vertical lines and filled boxes represent the statistical and systematic uncertainties, respectively.
The lower panels show the ratio of the distributions from the MC / \ppp data.   The grey band on the ratio plot represents the quadrature sum of the data statistical and systematic uncertainties.
}
\label{fig:resultPPComp}
\end{center}
\end{figure}

Results for the three photon plus two-jet observables (\xjjg, \ajjg, \drjj) are shown in this section for measurements in \ppp and in the three centrality intervals of Pb+Pb collisions.
First, to establish a baseline expectation for these observables in \ppp collisions, Figure~\ref{fig:resultPPComp} shows the three observables in \ppp collisions at 5.02~\TeV.
Figure~\ref{fig:resultPPComp}
also shows distributions obtained from the \ppp{} Monte Carlo generators \pythia-8, \jewel pp, and \MADGRAPH~\cite{Alwall:2014hca}.   A description of the \pythia-8 simulations was provided in Sec.~\ref{sec:dataMC}.     The \jewel (Jet Evolution With Energy Loss) framework includes an MC generator that models jet-quenching processes in QGP~\cite{Zapp:2008gi}, and \jewel pp is the result of running \jewel in \ppp collision mode, which is based on a modified version of \pythia-6.   In the \MADGRAPH case, the signal process, $\gamma + \text{jet}$, is simulated at next-to-leading-order
accuracy using the \textsc{MadGraph5\_aMC@NLO} v3.3.2 framework~\cite{Alwall:2014hca}.
The proton structure is modelled using the CT14nlo~\cite{Dulat:2015mca} parton
distribution function (PDF) set, accessed through the \textsc{LHAPDF6}
library~\cite{Buckley:2014ana}. To account for higher-order effects,
parton showering, and hadronisation, the matrix-element events are
subsequently interfaced with \textsc{Pythia} 8.307~\cite{Bierlich:2022pfr}.

All  generators capture the main trends for the three observables, with \MADGRAPH slightly over-predicting and \pythia-8 and \jewel pp slightly under-predicting the overall yield.     The generators also show more significant deviations from data in the tails of the \xjjg and \ajjg distributions.

\begin{figure}[hbtp!]
\begin{center}
\subfigure[]{\includegraphics[width=0.32\textwidth]{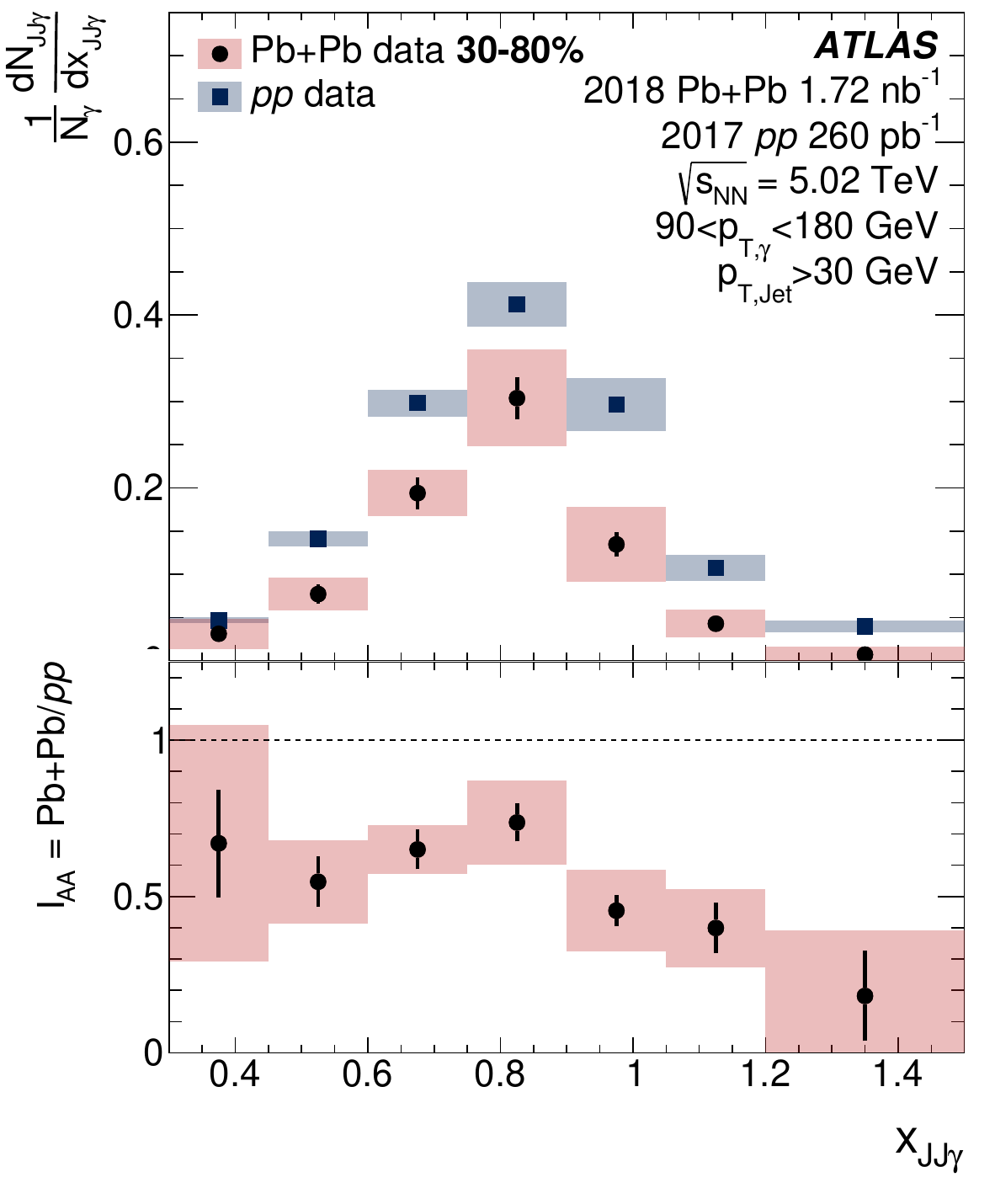} }
\subfigure[]{\includegraphics[width=0.32\textwidth]{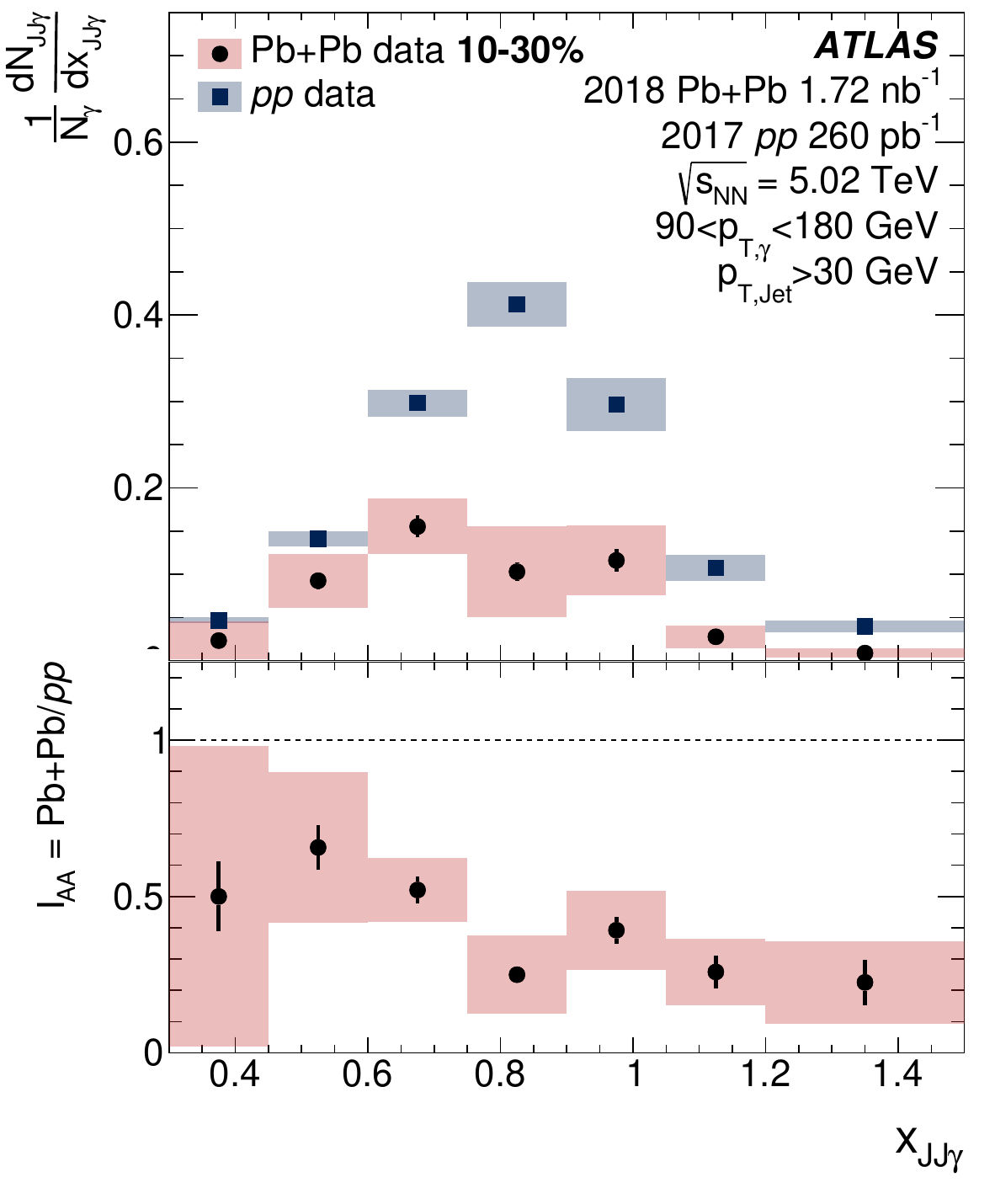}}
\subfigure[]{\includegraphics[width=0.32\textwidth]{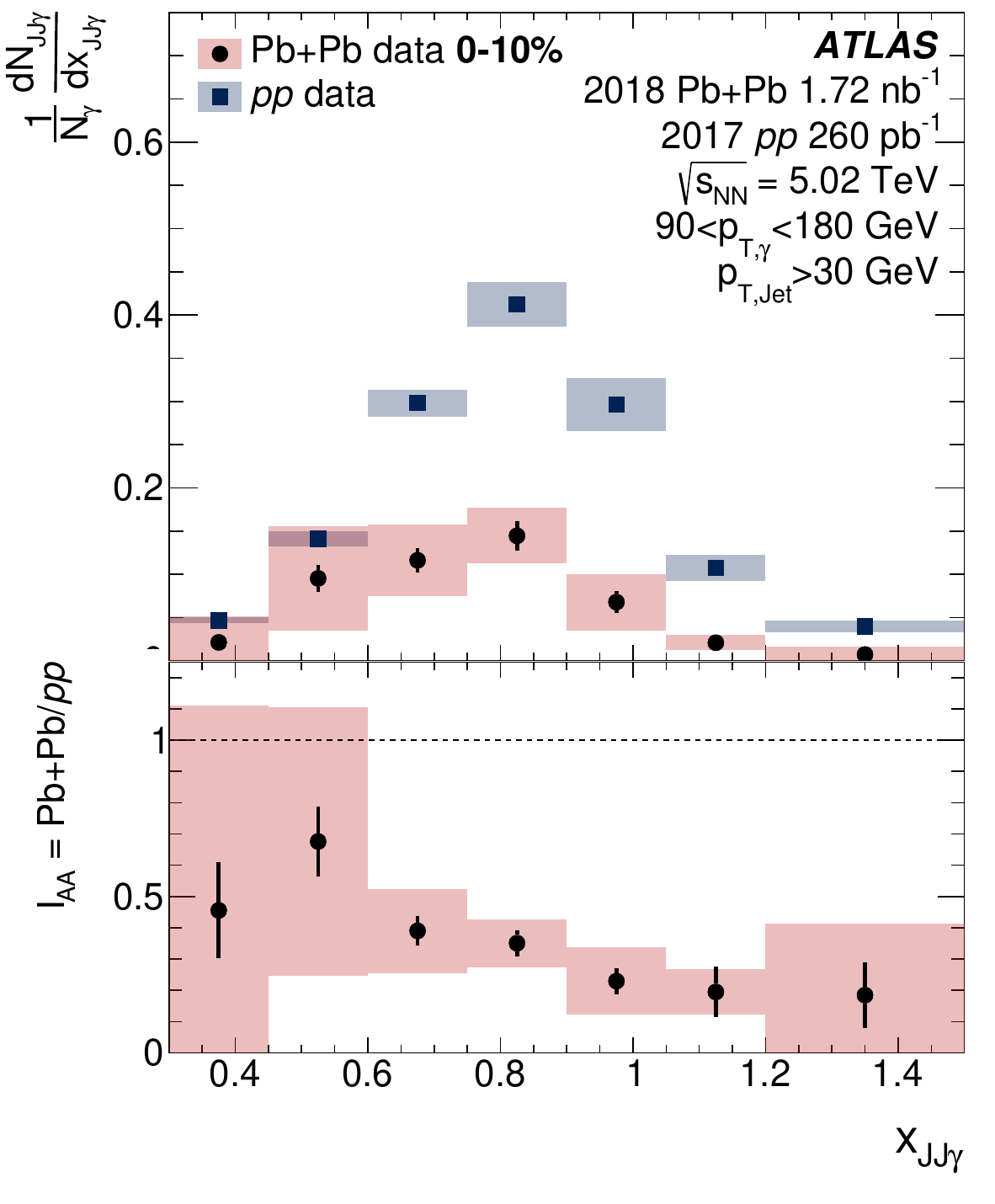}}
\caption{
Unfolded \xjjg distributions in \ppp and Pb+Pb collisions for the three centrality intervals in panels (a) 30--80\%, (b) 10--30\%, and (c) 0--10\%. The lower panels show \IAA, the ratio of distributions in Pb+Pb / \ppp collisions.
The vertical lines and filled boxes represent the statistical and systematic uncertainties, respectively.
The vertical lines and filled boxes represent the statistical and systematic uncertainties, respectively.The vertical lines and filled boxes represent the statistical and systematic uncertainties, respectively.}
\label{fig:resultXJJG}
\end{center}
\end{figure}

\begin{figure}[hbtp!]
\begin{center}
\subfigure[]{\includegraphics[width=0.32\textwidth]{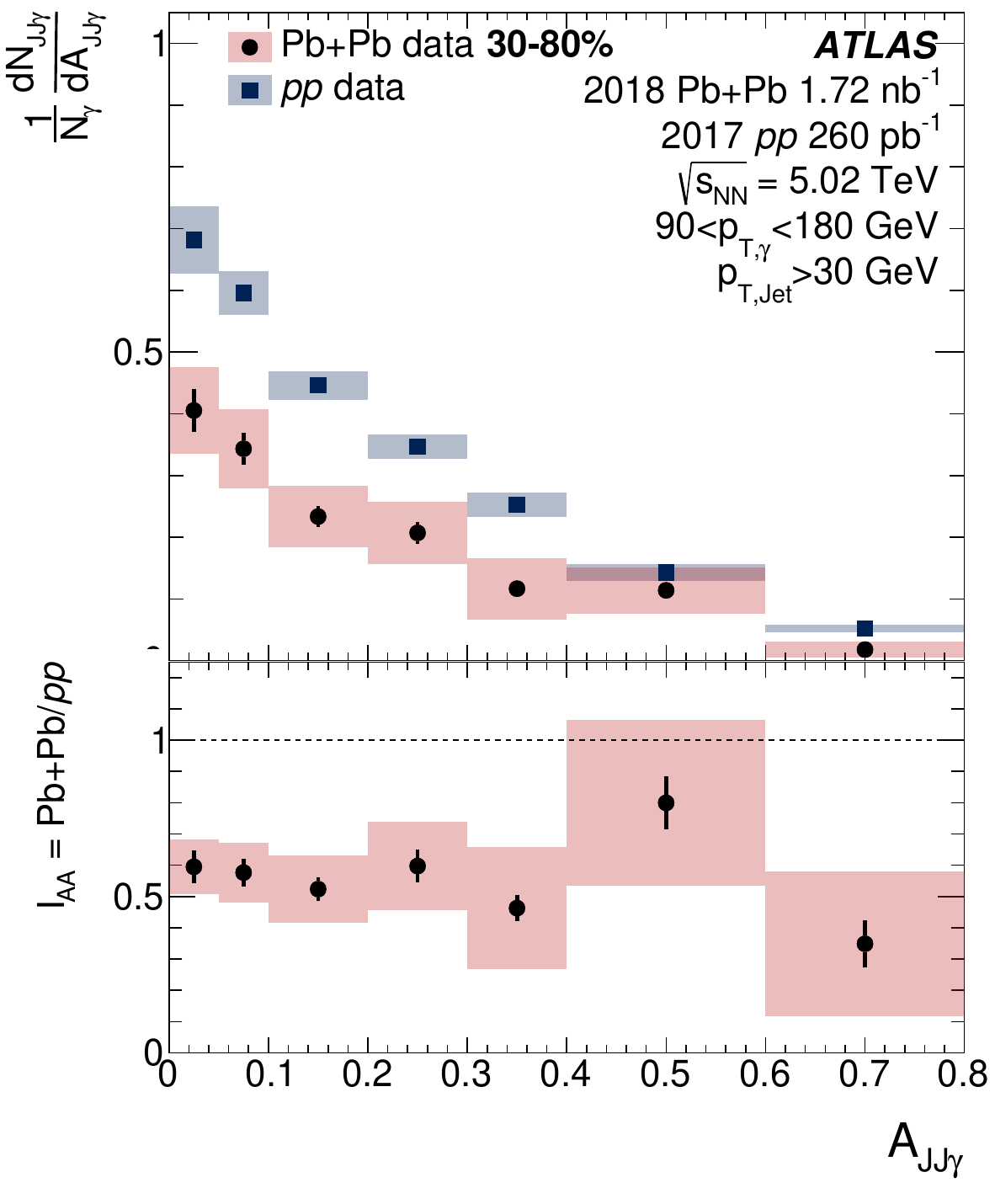} }
\subfigure[]{\includegraphics[width=0.32\textwidth]{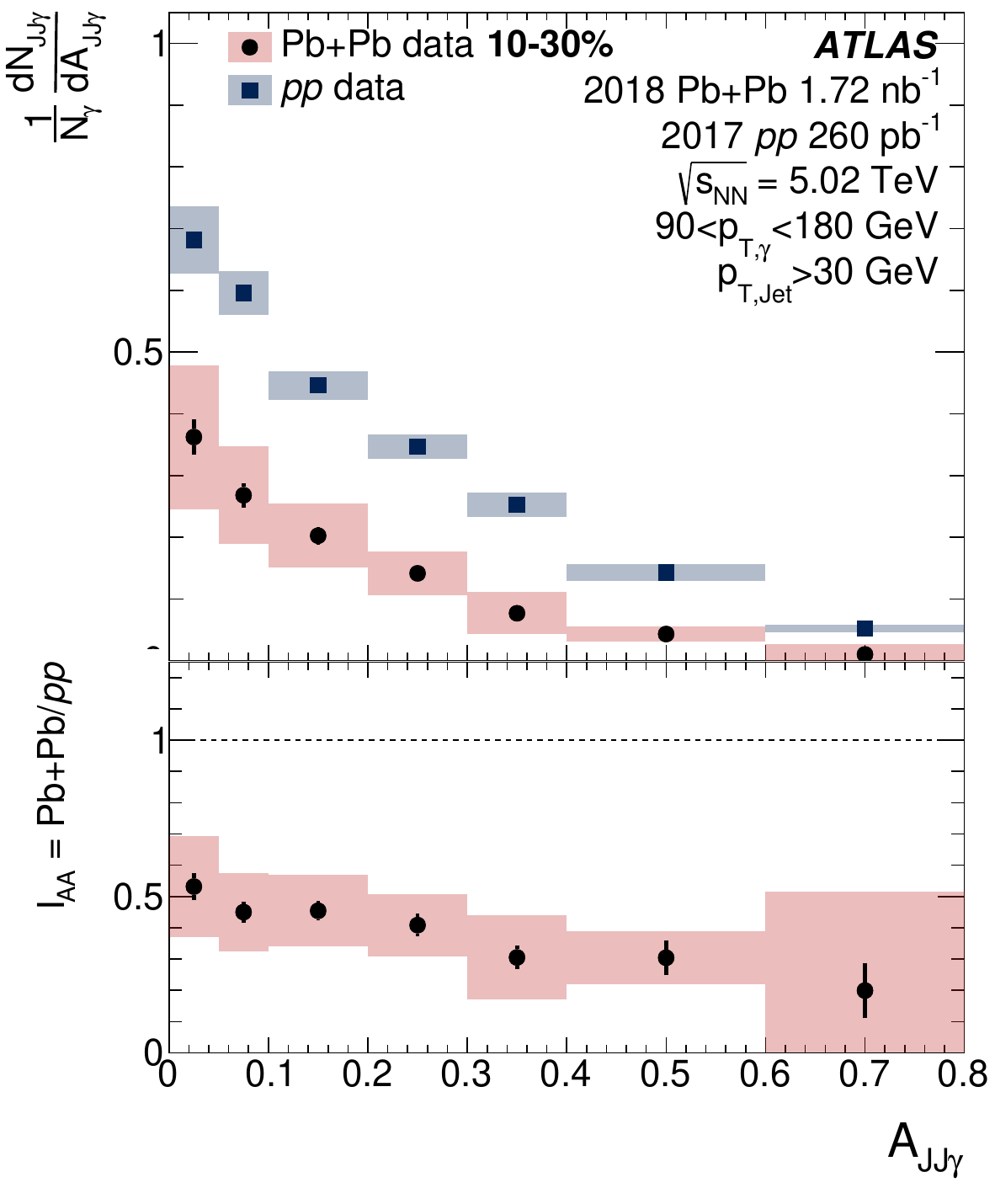}}
\subfigure[]{\includegraphics[width=0.32\textwidth]{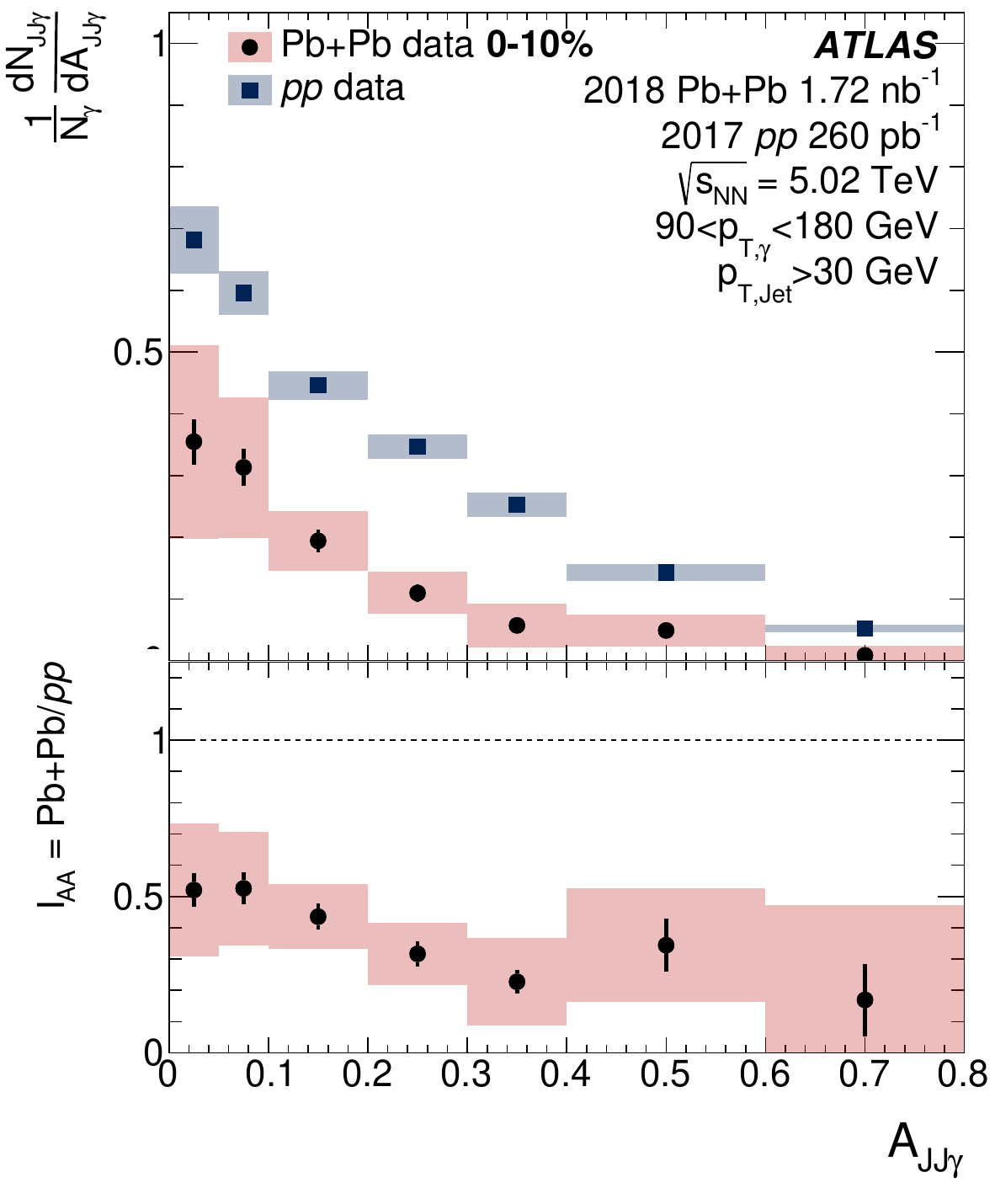}}
\caption{
Unfolded \ajjg
distributions in \ppp and Pb+Pb collisions for the three centrality intervals in panels (a) 30--80\%, (b) 10--30\%, and (c) 0--10\%.
The lower panels show \IAA, the ratio of distributions in Pb+Pb / \ppp collisions.   The vertical lines and filled boxes represent the statistical and systematic uncertainties, respectively.
}
\label{fig:resultAJJG}
\end{center}
\end{figure}

\begin{figure}[hbtp!]
\begin{center}
\subfigure[]{\includegraphics[width=0.32\textwidth]{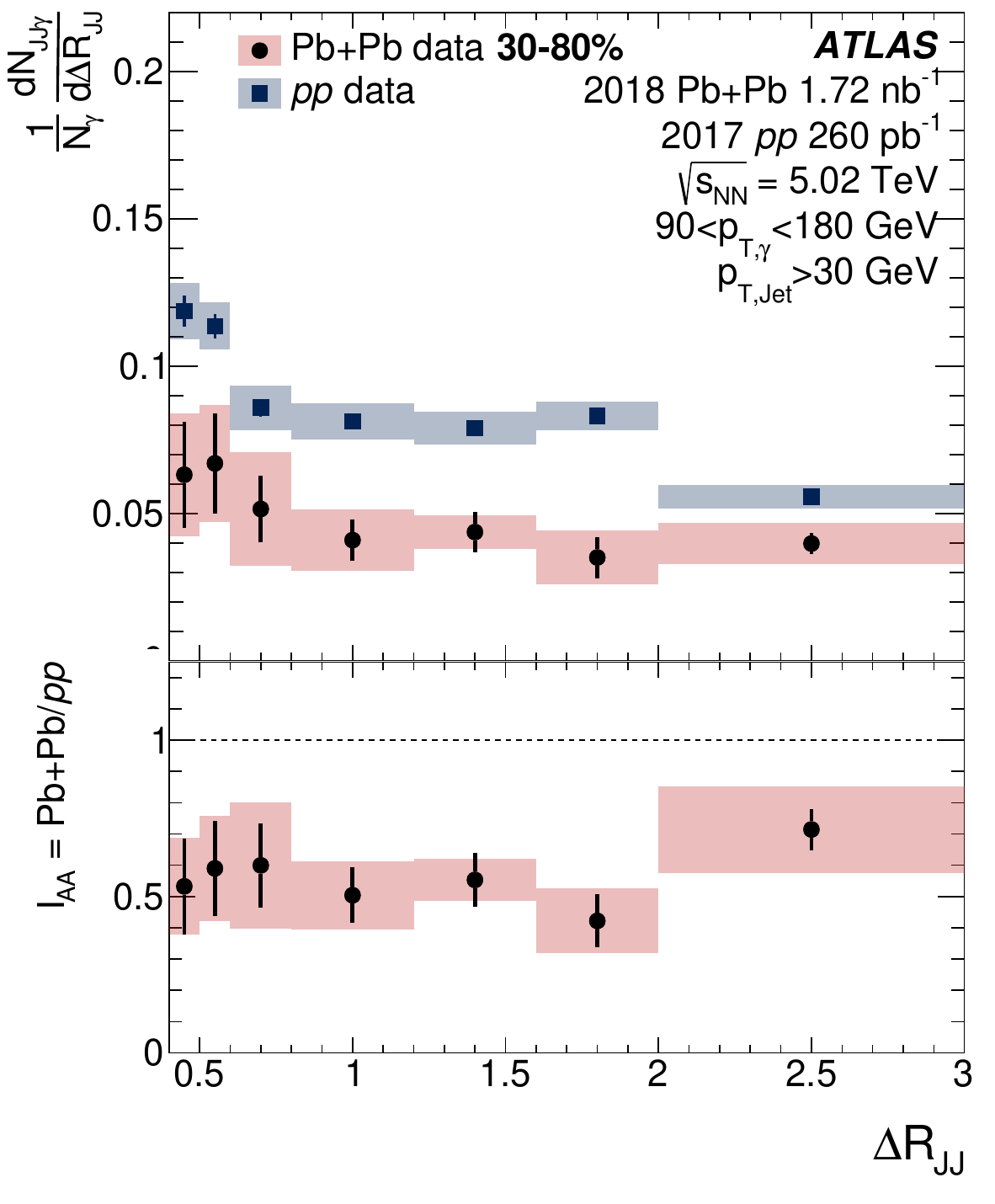} }
\subfigure[]{\includegraphics[width=0.32\textwidth]{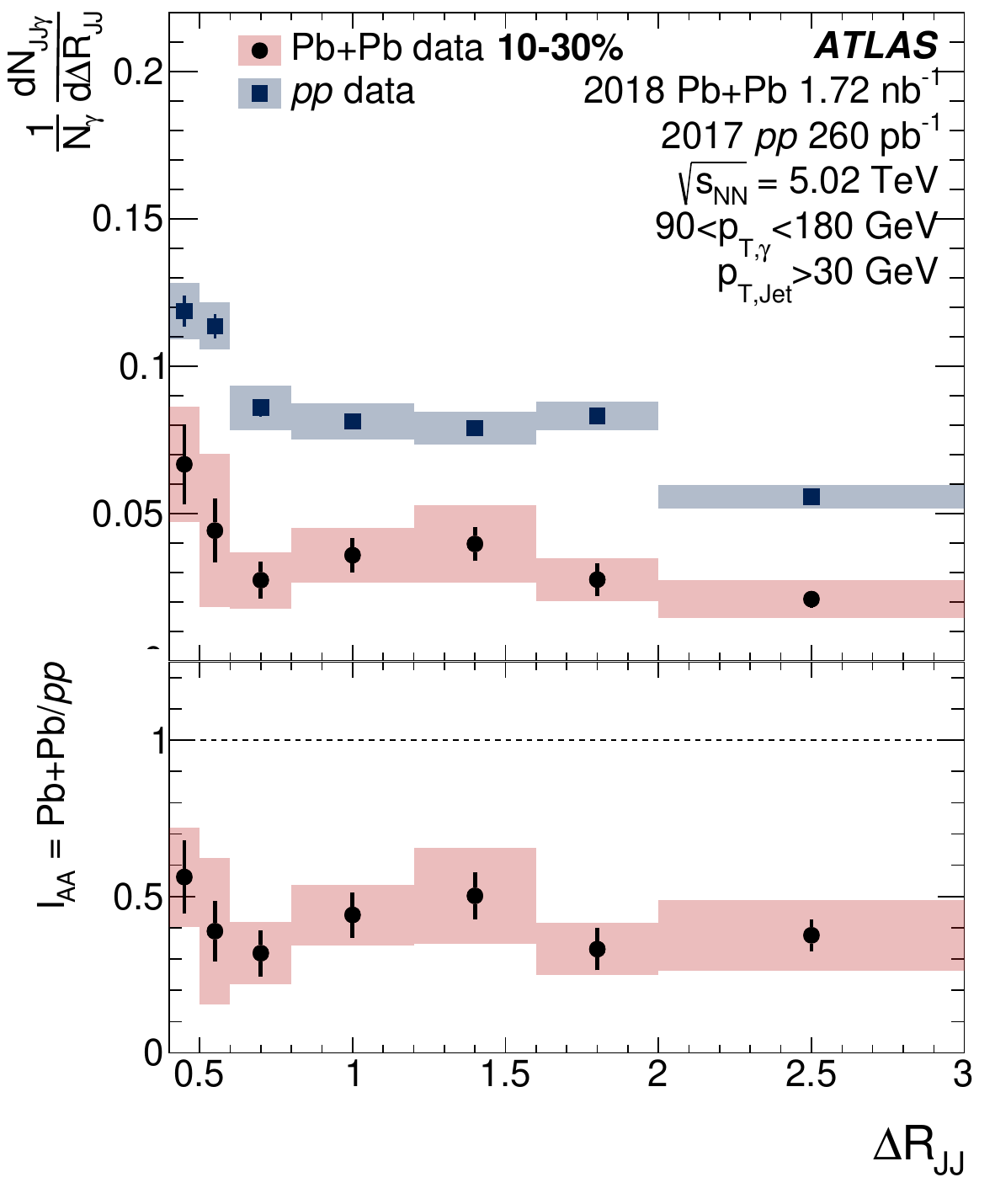}}
\subfigure[]{\includegraphics[width=0.32\textwidth]{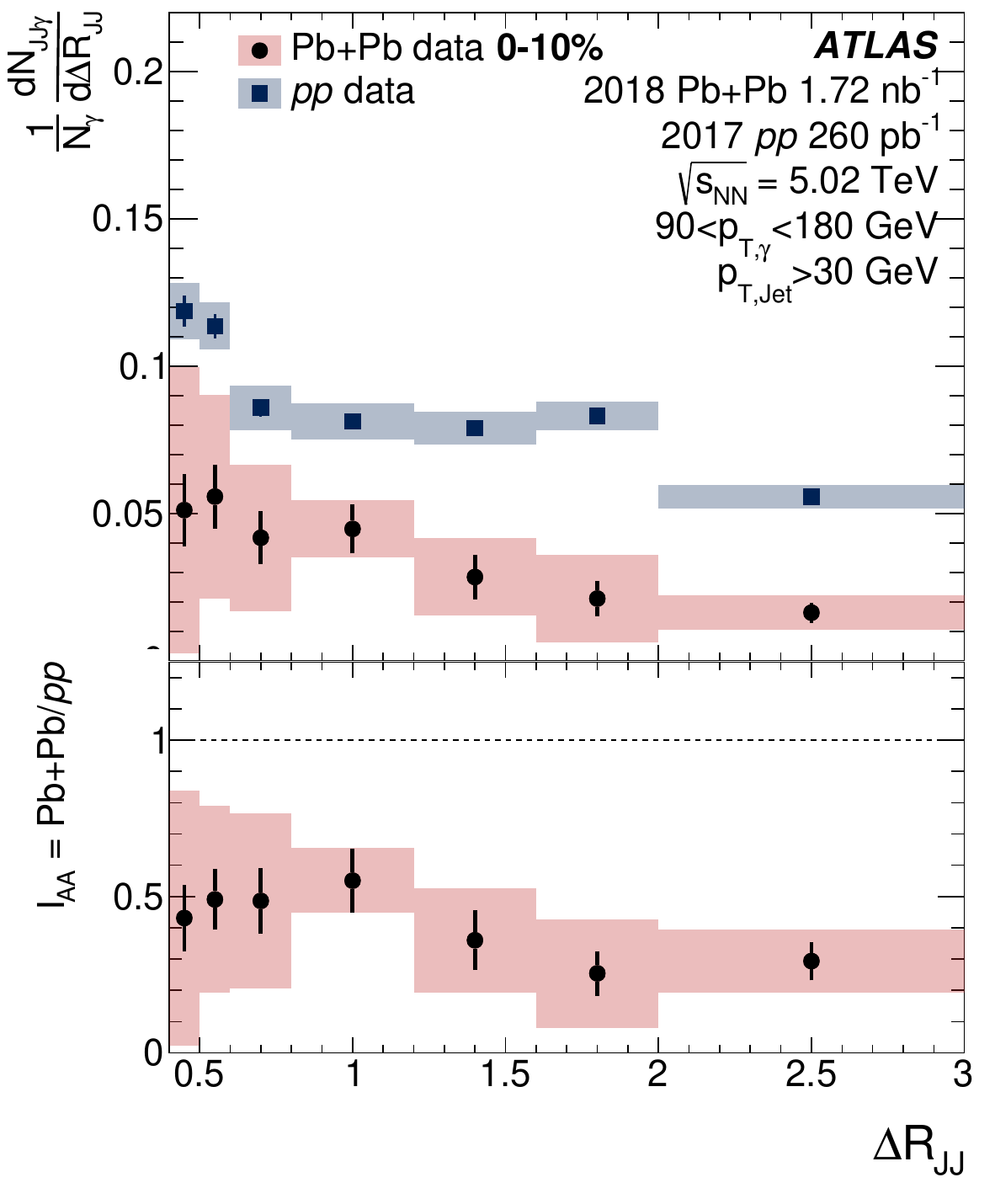}}
\caption{
Unfolded \drjj distributions in \ppp and Pb+Pb collisions for the three centrality intervals in panels (a) 30--80\%, (b) 10--30\%, and (c) 0--10\%.
The lower panels show \IAA, the ratio of distributions in Pb+Pb / \ppp collisions.
The vertical lines and filled boxes represent the statistical and systematic uncertainties, respectively.
}
\label{fig:resultDRJJ}
\end{center}
\end{figure}

Figures~\ref{fig:resultXJJG}, \ref{fig:resultAJJG} and \ref{fig:resultDRJJ} show the Pb+Pb results in the 30--80\%, 10--30\%, and 0--10\% centrality intervals for the three observables.
The upper panel of each figure shows the fully corrected and unfolded distributions in Pb+Pb collisions for a given centrality interval and the common \ppp data.
The ratio of distributions in Pb+Pb collisions divided by those in \ppp collisions, \IAA, is shown in the lower panels.

For all three observables in all centrality ranges, there is a significant overall suppression in Pb+Pb compared with \ppp collisions (\IAA $<1$).    The suppression is the smallest for the most peripheral (30--80\%) collisions, as expected, since the QGP created is smaller in volume and of lower temperature~\cite{Habich:2014jna}.     In the most central collisions (0--10\%), the \IAA is smaller for larger values of \xjjg, \ajjg, and \drjj.

\clearpage
\newpage

The experimental results in the Pb+Pb
0--10\% centrality interval are compared with three theoretical jet quenching models referred to as \jewel, \jetscape, and \lbt.
The \jewel MC generator modifies the parton shower evolution via interactions with a medium, modelled as a two-dimensional outwardly expanding QGP\@.   \jewel does not incorporate colour-coherence effects.
Version 2.2.0 of \jewel is publicly available following the initial release detailed in Ref.~\cite{Zapp:2013vla}.
\jewel was run using the photon+jet hard-scattering process~\cite{KunnawalkamElayavalli:2016ttl} that includes only direct (leading-order) photon production.
The photon and jet kinematic selections used in the data analysis are applied and the observables are calculated.

\jetscape (Jet Energy-loss Tomography with a Statistically and Computationally Advanced Program Envelope) is a framework that enables different modules to be plugged in for modelling the initial collision geometry, different jet kinematics, and different physics processes of jet quenching~\cite{JETSCAPE:2023hqn}.    The \jetscape authors have provided calculation results for \ppp (based on \pythia-8) and 0--10\% Pb+Pb collisions using the plug-in modules \matter (Modular All Twist
Transverse-scattering Elastic-drag and Radiation) + \lbt (Linearized Boltzmann Transport) with the \jetscape v3.5 AA22 tune, identical to that used in Refs.~\cite{JETSCAPE:2022jer,JETSCAPE:2022hcb,JETSCAPE:2023hqn}.
The \matter module handles the medium interactions at high virtuality (above some $Q_{0}$ parameter) where modified coherence effects diminish medium-induced radiation, and a lower virtuality phase is modelled using the Linear Boltzmann Transport model (\lbt)~\cite{Luo:2018pto}, where parton splittings are fully resolved by the medium as they undergo multiple-scattering-induced energy loss.

\lbt calculations were also run separately (without the \matter module) using \MADGRAPH~\cite{Alwall:2014hca} for the \ppp baseline and for initial hard scatterings as input to the QGP medium propagation.

\begin{figure}[hbtp!]
\begin{center}
\subfigure[]{\includegraphics[width=0.32\textwidth]{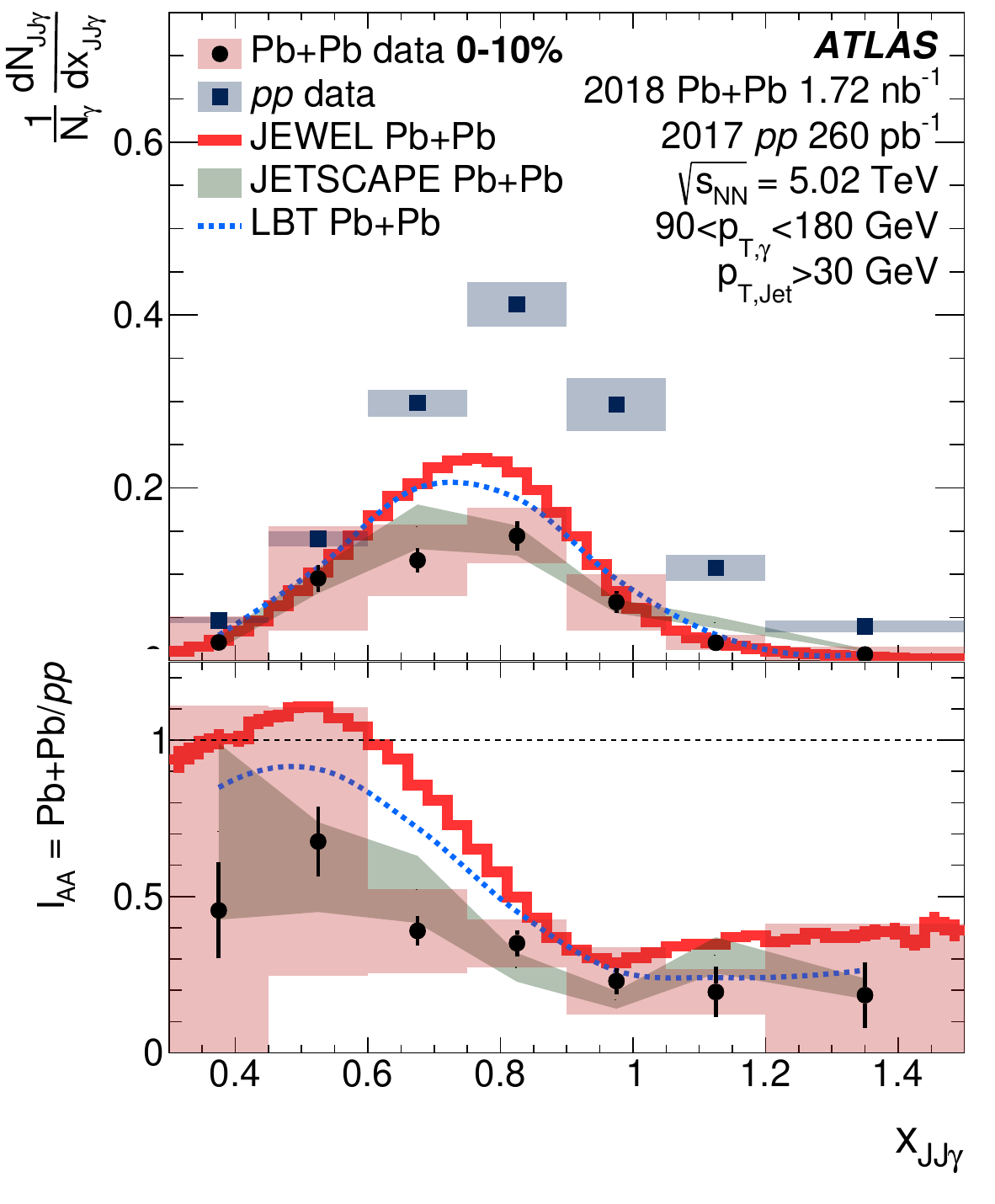}}
\subfigure[]{\includegraphics[width=0.32\textwidth]{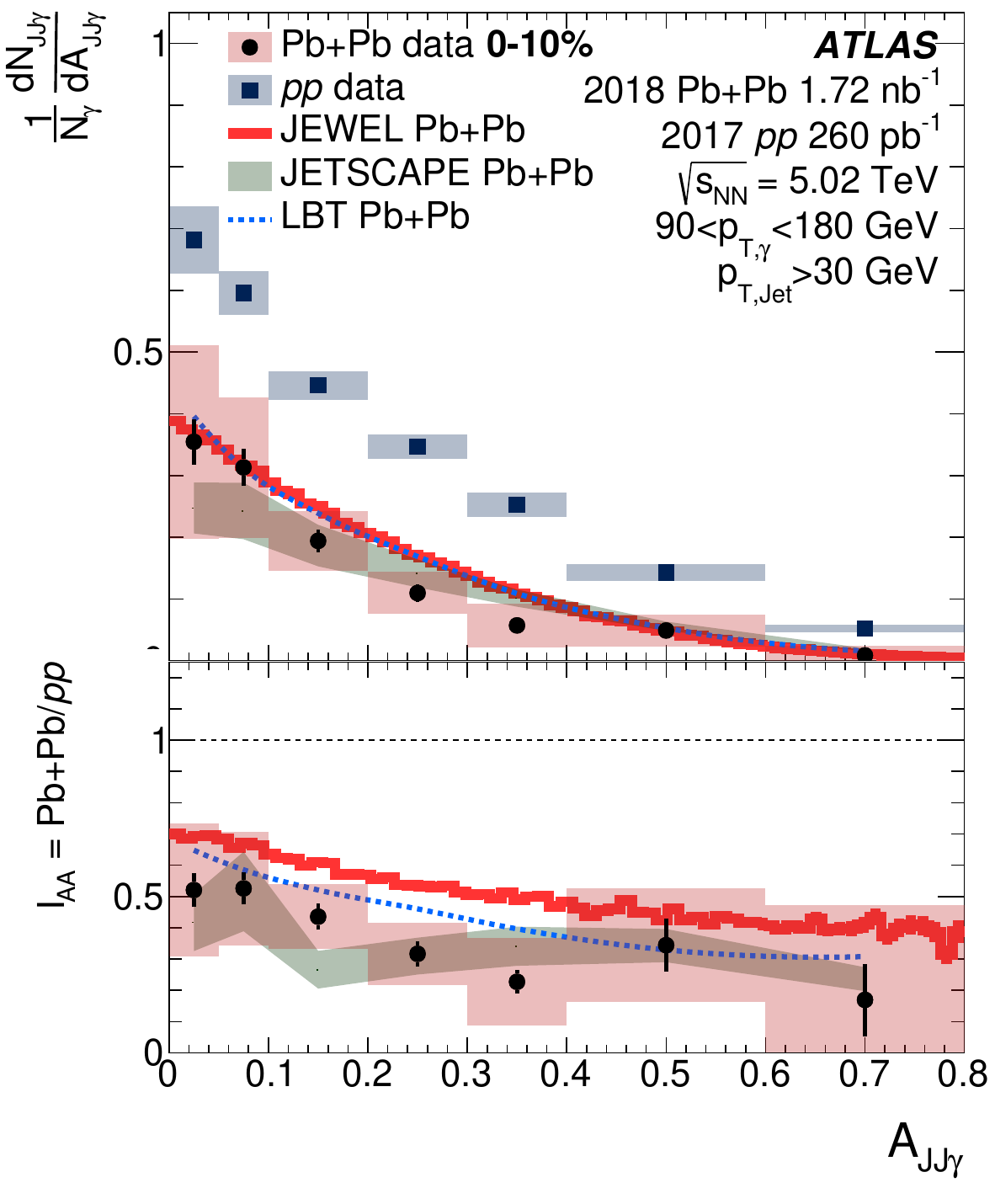}}
\subfigure[]{\includegraphics[width=0.32\textwidth]{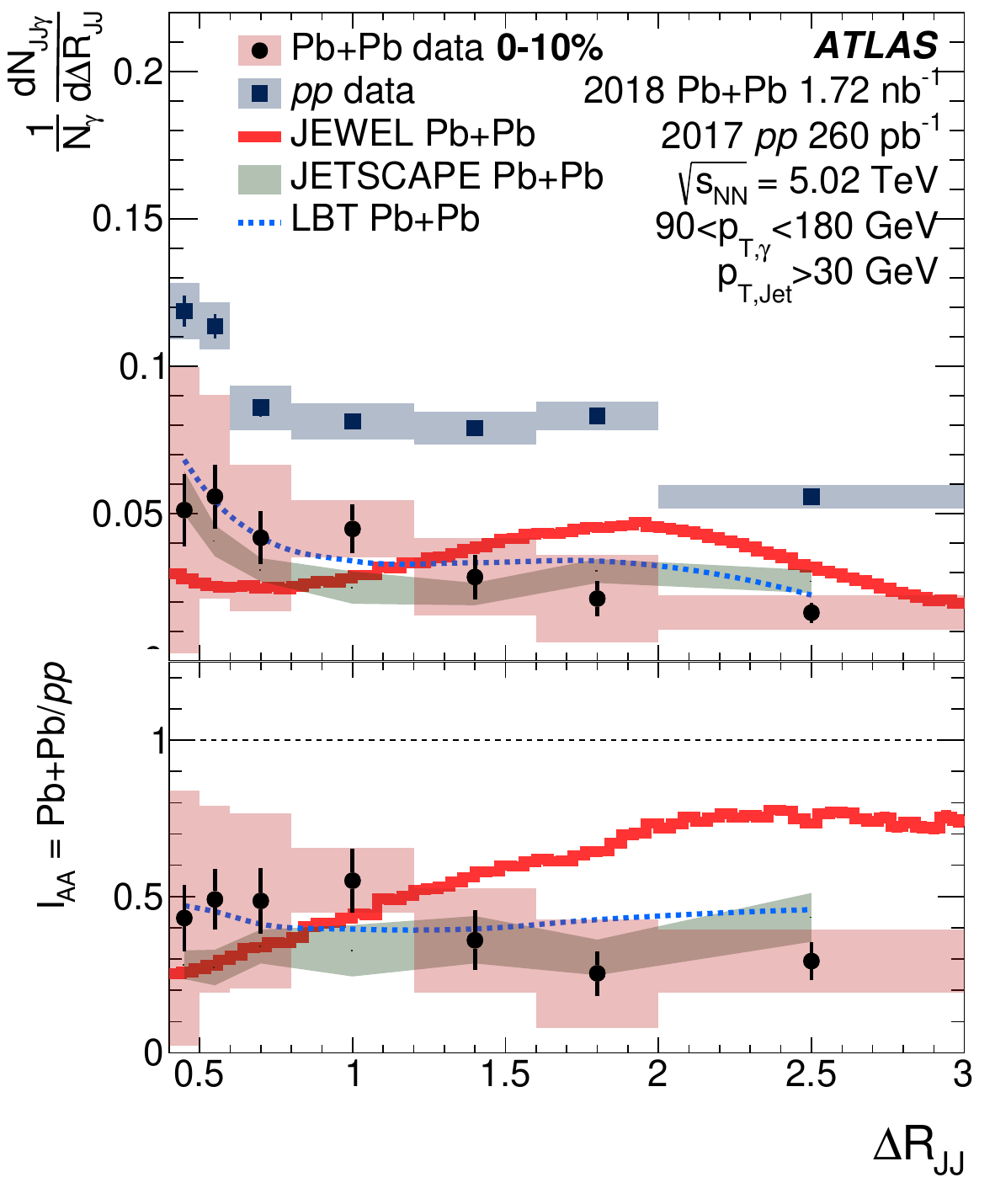}}
\caption{
Unfolded distribution in \ppp and 0--10\% Pb+Pb collisions for (a) \xjjg, (b) \ajjg, and (c) \drjj. The lower panels show \IAA, the ratio of per-photon yield in Pb+Pb / \ppp collisions.
The vertical lines and filled boxes represent the statistical and systematic uncertainties, respectively.
Also shown are theoretical calculations from \jewel, \jetscape, and \lbt in Pb+Pb collisions and for the \IAA (lower panels).
}
\label{fig:super10}
\end{center}
\end{figure}

Figure~\ref{fig:super10} shows the three theoretical models compared with experimental data for all three observables.    The \xjjg distribution is qualitatively described by all models, with a substantially larger suppression, i.e. smaller \IAA, at \xjjg $> 0.8$.    Similarly, the \ajjg distribution is well described by  all models, with the \jewel result slightly over-predicting the \IAA.    Lastly, the \drjj distribution is described by both \jetscape and \lbt, but has the completely  opposite trend in \jewel.   The data indicates somewhat less suppression of dijet events with larger angular separation, whereas \jewel has a substantial suppression of dijet events with small angular separation.

\begin{figure}[hbtp!]
\begin{center}
\subfigure[]{\includegraphics[width=0.32\textwidth]{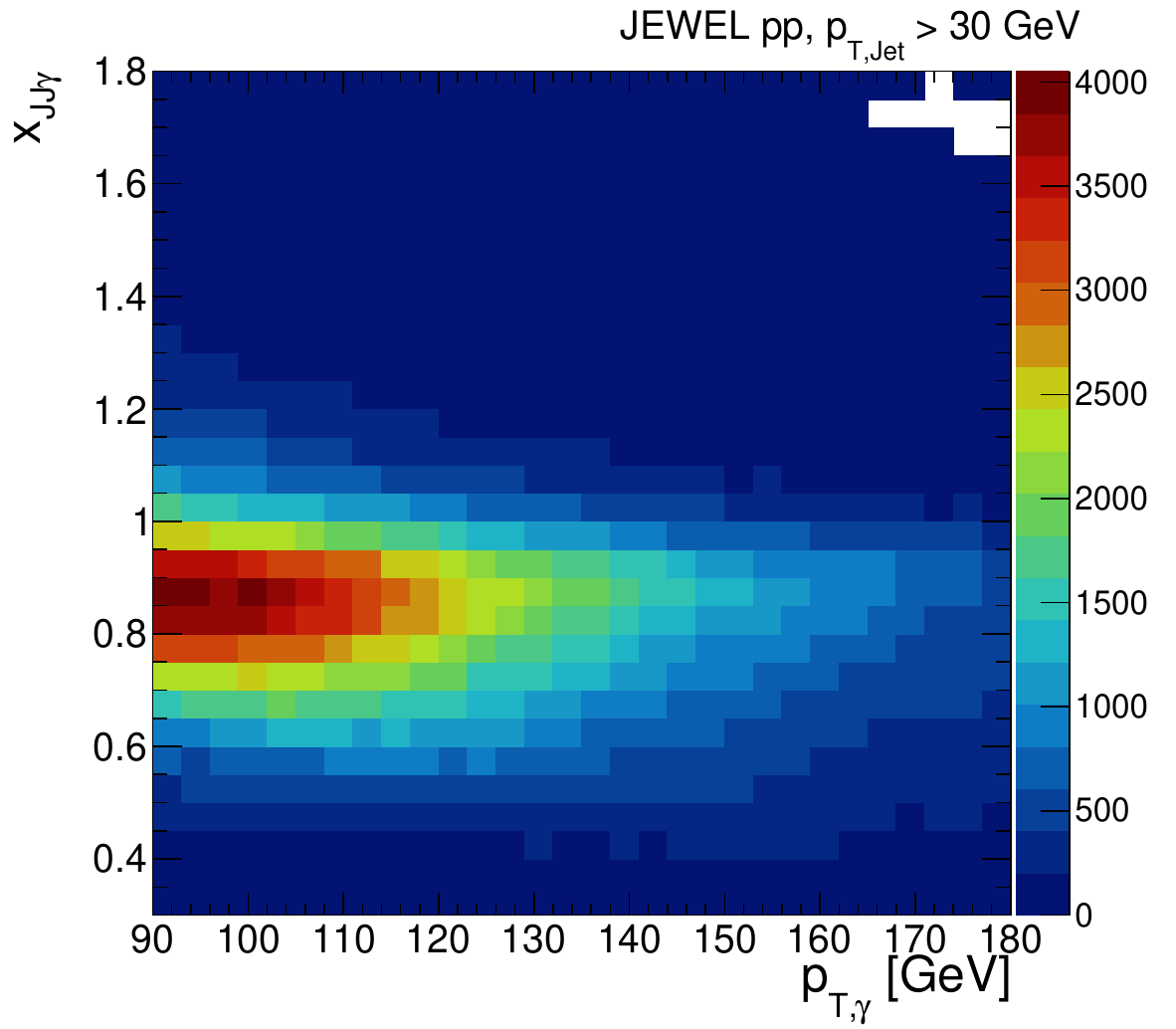}}
\subfigure[]{\includegraphics[width=0.32\textwidth]{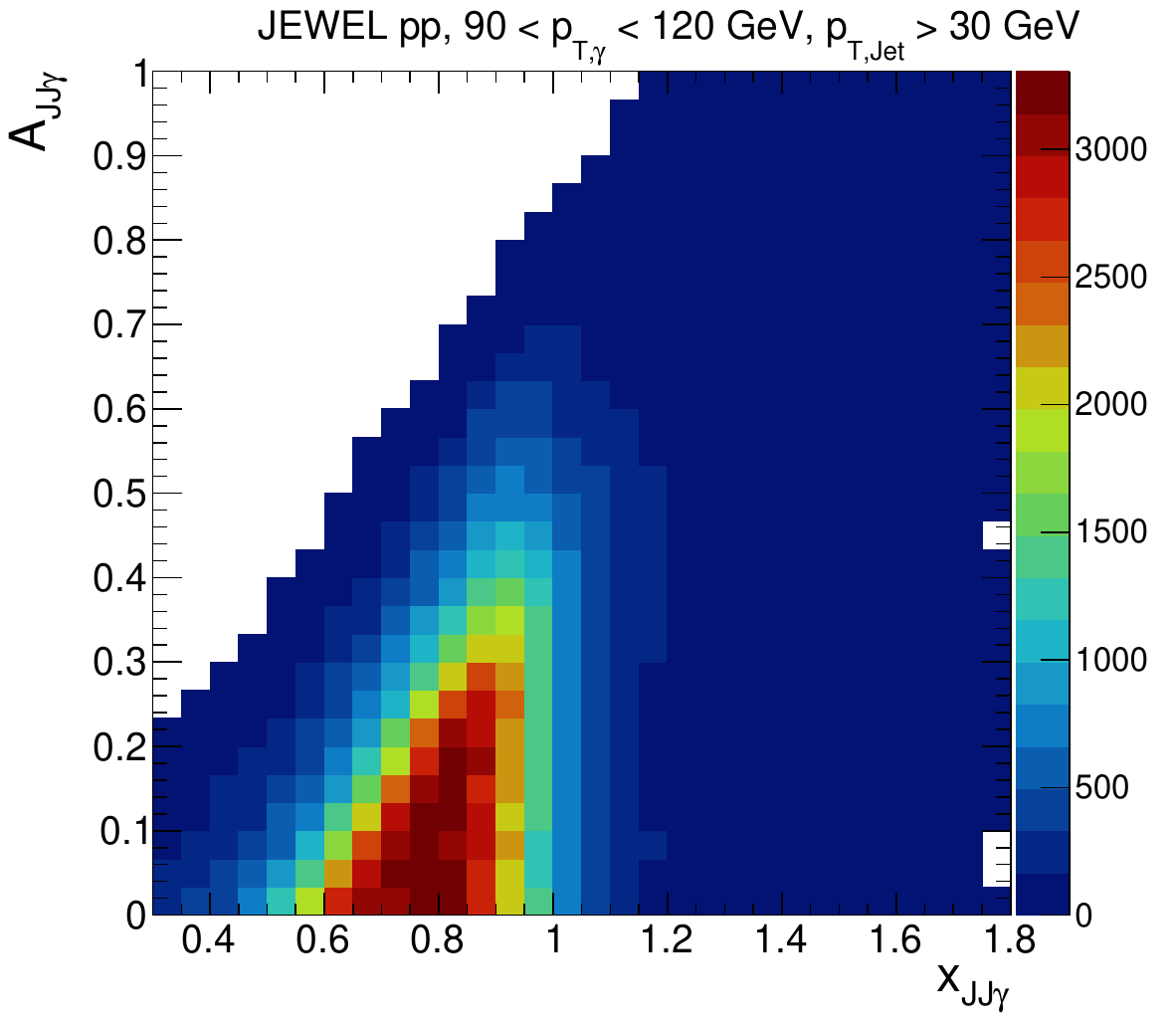}}
\subfigure[]{\includegraphics[width=0.32\textwidth]{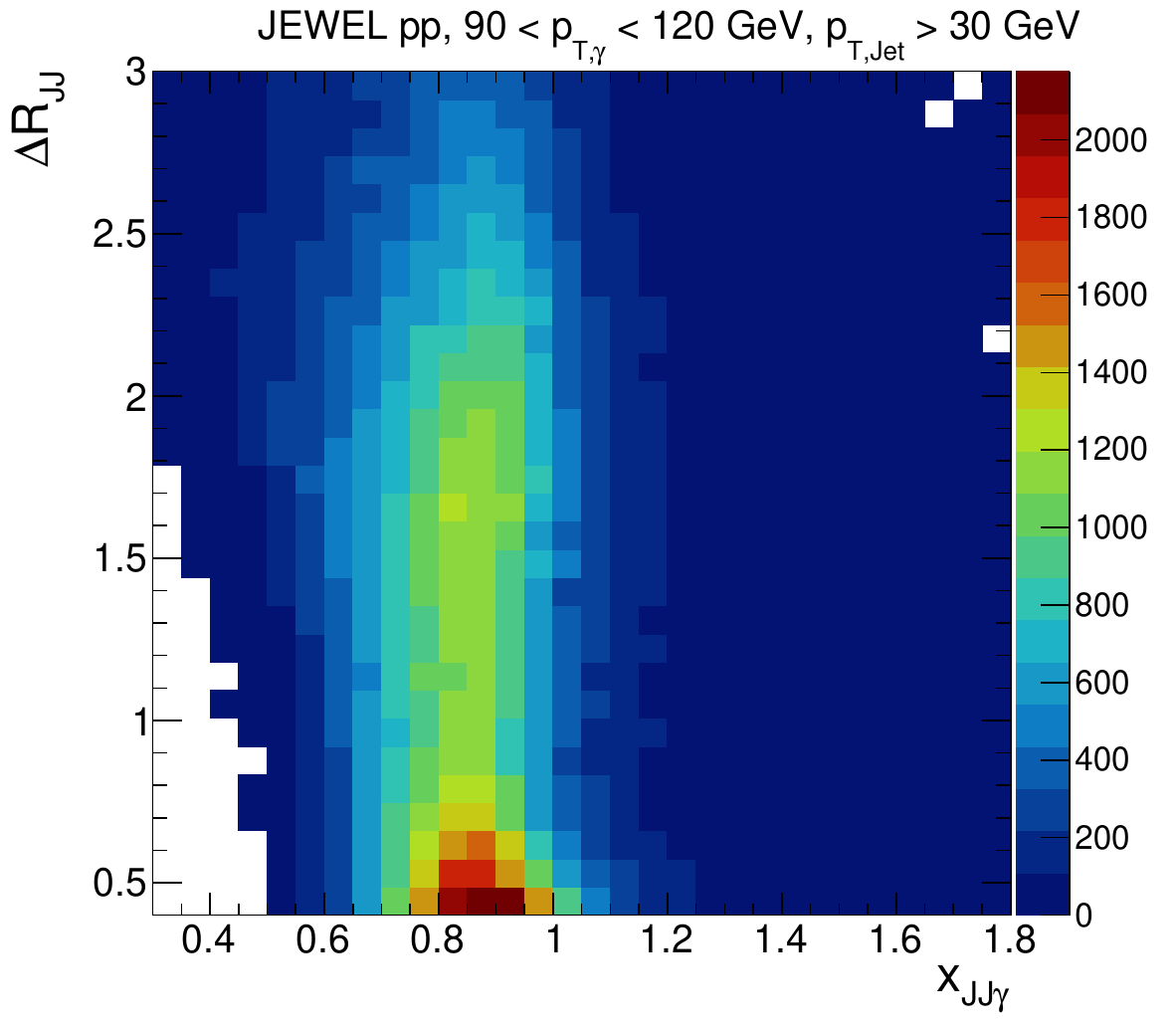}}
\caption{
Monte Carlo generator \jewel ~\ppp results illustrating the kinematic correlations between the observables in the absence of quenching effects.   Panel (a) shows the \xjjg as a function of \ptg.   Panel (b) shows \ajjg as a function of \xjjg.   Panel (c) shows \drjj as a function of \xjjg.     The distributions are generated with $90 < \ptg < 120$~GeV and $\ptj > 30$~GeV, with
all other
kinematic selections identical to those used in this analysis.
}
\label{fig:kinematics}
\end{center}
\end{figure}

Figure~\ref{fig:kinematics} shows results from the \jewel{} \ppp{} Monte Carlo generator for the correlation between the different observables with panel (a) showing \xjjg versus \ptg, panel (b) showing \ajjg versus \xjjg, and (c) showing \drjj versus \xjjg.   It is notable that at the lowest \ptg = 90~\GeV, the acceptance in \ajjg is limited to values predominantly below 0.3, and larger \ajjg values are only populated from higher \ptg events.    For \xjjg and \drjj there are no such strong kinematic correlations.  The MC generator calculations have the identical kinematic selections as the data.   It is notable that all generators show a suppression at \xjjg $\approx$ 1, which automatically suppresses the yield for \ajjg $>$ 0.3.

\FloatBarrier

\section{Conclusion}

This paper  presents results of a measurement by ATLAS of photon plus two-jet observables in \ppp and Pb+Pb collisions at 5.02~\TeV collected in 2017 and 2018, respectively, to study the impact of medium-parton interactions on nearby quark and gluon probes of the QGP\@. Photons are selected with $90 < \ptg < 180$~\GeV and $R=0.2$ jets are selected with $\pt > 30$~\GeV.
The unfolded distributions are measured for three observables, \xjjg, \ajjg, and \drjj, in \ppp data and three Pb+Pb centrality intervals (30--80\%, 10--30\%, and 0--10\%). Ratios of the distributions in each Pb+Pb centrality interval to those in \ppp collisions (\IAA) are also measured. Significant suppression ($\IAA < 1$) is observed in all cases, with the suppression larger in the most central relative to the most peripheral centrality intervals.
Monte Carlo calculations from the jet quenching models \jewel, \jetscape, and \lbt are compared with the experimental data.
These measurements provide new constraints on the microscopic processes of energy loss in the QGP, and further tests of the underlying assumptions of theoretical models will help refine their description of these processes.

\FloatBarrier

\section*{Acknowledgements}
%

%
%

%
%

We thank CERN for the very successful operation of the LHC and its injectors, as well as the support staff at
CERN and at our institutions worldwide without whom ATLAS could not be operated efficiently.

The crucial computing support from all WLCG partners is acknowledged gratefully, in particular from CERN, the ATLAS Tier-1 facilities at TRIUMF/SFU (Canada), NDGF (Denmark, Norway, Sweden), CC-IN2P3 (France), KIT/GridKA (Germany), INFN-CNAF (Italy), NL-T1 (Netherlands), PIC (Spain), RAL (UK) and BNL (USA), the Tier-2 facilities worldwide and large non-WLCG resource providers. Major contributors of computing resources are listed in Ref.~\cite{ATL-SOFT-PUB-2026-001}.

We gratefully acknowledge the support of ANPCyT, Argentina; YerPhI, Armenia; ARC, Australia; BMWFW and FWF, Austria; ANAS, Azerbaijan; CNPq and FAPESP, Brazil; NSERC, NRC and CFI, Canada; CERN; ANID, Chile; CAS, MOST and NSFC, China; Minciencias, Colombia; MEYS CR, Czech Republic; DNRF and DNSRC, Denmark; IN2P3-CNRS and CEA-DRF/IRFU, France; SRNSFG, Georgia; BMFTR, HGF and MPG, Germany; GSRI, Greece; RGC and Hong Kong SAR, China; ICHEP and Academy of Sciences and Humanities, Israel; INFN, Italy; MEXT and JSPS, Japan; CNRST, Morocco; NWO, Netherlands; RCN, Norway; MNiSW, Poland; FCT, Portugal; MNE/IFA, Romania; MSTDI, Serbia; MSSR, Slovakia; ARIS and MVZI, Slovenia; DSI/NRF, South Africa; MICIU/AEI, Spain; SRC and Wallenberg Foundation, Sweden; SERI, SNSF and Cantons of Bern and Geneva, Switzerland; NSTC, Taipei; TENMAK, T\"urkiye; STFC/UKRI, United Kingdom; DOE and NSF, United States of America.

Individual groups and members have received support from BCKDF, CANARIE, CRC and DRAC, Canada; CERN-CZ, FORTE and PRIMUS, Czech Republic; COST, ERC, ERDF, Horizon 2020, ICSC-NextGenerationEU and Marie Sk{\l}odowska-Curie Actions, European Union; Investissements d'Avenir Labex, Investissements d'Avenir Idex and ANR, France; DFG and AvH Foundation, Germany; Herakleitos, Thales and Aristeia programmes co-financed by EU-ESF and the Greek NSRF, Greece; BSF-NSF and MINERVA, Israel; NCN and NAWA, Poland; La Caixa Banking Foundation, CERCA and AGAUR programs from Generalitat de Catalunya and PROMETEO and GenT Programmes Generalitat Valenciana, Spain; G\"{o}ran Gustafssons Stiftelse, Sweden; The Royal Society and Leverhulme Trust, United Kingdom; Eric and Wendy Schmidt Fund for Strategic Innovation, United States of America.

In addition, individual members wish to acknowledge support from Armenia: Yerevan Physics Institute (FAPERJ); CERN: European Organization for Nuclear Research (CERN DOCT); Chile: Agencia Nacional de Investigaci\'on y Desarrollo (ANID FONDECYT reg. 1230987, FONDECYT 1230812, FONDECYT 1240864, Fondecyt 3240661, Fondecyt Regular 1240721); China: Chinese Ministry of Science and Technology (MOST-2023YFA1605700, MOST-2023YFA1609300), National Natural Science Foundation of China (NSFC - 12175119, NSFC 12275265); Czech Republic: Czech Science Foundation (GACR - 24-11373S), Ministry of Education Youth and Sports (ERC-CZ-LL2327, FORTE CZ.02.01.01/00/22\_008/0004632), PRIMUS Research Programme (PRIMUS/21/SCI/017); EU: H2020 European Research Council (ERC - 101002463); European Union: European Research Council (BARD No. 101116429, ERC - 948254, ERC 101089007), European Regional Development Fund (HE COFUND GA No.101081355, ERDF), Horizon 2020 Framework Programme (MUCCA - CHIST-ERA-19-XAI-00), European Union, Future Artificial Intelligence Research (FAIR-NextGenerationEU PE00000013), Horizon 2020 (EuroHPC - EHPC-DEV-2024D11-051), Italian Center for High Performance Computing, Big Data and Quantum Computing (ICSC, NextGenerationEU), Marie Sklodowska-Curie Actions (ERC Horizon-2020, GA 956086, GAP-101168829); France: Agence Nationale de la Recherche (ANR-21-CE31-0013, ANR-21-CE31-0022, ANR-22-EDIR-0002, ANR-24-CE31-0504-01); Germany: Baden-Württemberg Stiftung (BW Stiftung-Postdoc Eliteprogramme), Deutsche Forschungsgemeinschaft (DFG - 469666862, DFG - CR 312/5-2); China: Research Grants Council (GRF); Italy: Istituto Nazionale di Fisica Nucleare (ICSC, NextGenerationEU), Ministero dell'Università e della Ricerca (NextGenEU 153D23001490006 M4C2.1.1, NextGenEU I53D23000820006 M4C2.1.1, NextGenEU I53D23001490006 M4C2.1.1, SOE2024\_0000023); Japan: Japan Society for the Promotion of Science (JSPS KAKENHI JP22H01227, JSPS KAKENHI JP22H04944, JSPS KAKENHI JP22KK0227, JSPS KAKENHI JP24K23939); Norway: Research Council of Norway (RCN-314472); Poland: Ministry of Science and Higher Education (IDUB AGH, POB8, D4 no 9722), Polish National Science Centre (NCN 2021/42/E/ST2/00350, NCN OPUS 2023/51/B/ST2/02507, NCN OPUS nr 2022/47/B/ST2/03059, NCN UMO-2019/34/E/ST2/00393, UMO-2022/47/O/ST2/00148, UMO-2023/49/B/ST2/04085, UMO-2023/51/B/ST2/00920, UMO-2024/53/N/ST2/00869); Portugal: Foundation for Science and Technology (FCT); Spain: Agència de Gestió d'Ajuts Universitaris i de Recerca. (AGAUR - 2023 BP 00141), Generalitat Valenciana (ASFAE/2022/008), Ministry of Science and Innovation (MCIN \& NextGenEU PCI2022-135018-2, MICIN \& FEDER PID2021-125273NB, RYC2019-028510-I, RYC2020-030254-I, RYC2021-031273-I, RYC2022-038164-I), Ministerio de Ciencia, Innovación y Universidades/Agencia Estatal de Investigaci\'on (EU NextGenerationEU (PRTR-C17.I1), PID2022-142604OB-C22); Sweden: Carl Trygger Foundation (Carl Trygger Foundation CTS 22:2312), Swedish Research Council (Swedish Research Council 2023-04654, VR 2021-03651, VR 2022-03845, VR 2022-04683, VR 2023-03403, VR 2024-05451), Knut and Alice Wallenberg Foundation (KAW 2018.0458, KAW 2022.0358, KAW 2023.0366); Switzerland: Swiss National Science Foundation (SNSF - PCEFP2\_194658); United Kingdom: Leverhulme Trust (Leverhulme Trust RPG-2020-004), Royal Society (NIF-R1-231091); United States of America: U.S. Department of Energy (ECA DE-AC02-76SF00515), Neubauer Family Foundation.

%
%


%
%
%
%
%
%
%

%
%
%
%
%
%

%
%

%
%
%
\printbibliography
\clearpage
\input{atlas_authlist}
%

%
%

%

%
%
%
%

\end{document}

%% file: atlas_authlist.tex
 
\begin{flushleft}
\hypersetup{urlcolor=black}
{\Large The ATLAS Collaboration}

\bigskip

\AtlasOrcid[0000-0002-6665-4934]{G.~Aad}$^\textrm{\scriptsize 104}$,
\AtlasOrcid[0000-0001-7616-1554]{E.~Aakvaag}$^\textrm{\scriptsize 17}$,
\AtlasOrcid[0000-0002-5888-2734]{B.~Abbott}$^\textrm{\scriptsize 122}$,
\AtlasOrcid[0000-0002-0287-5869]{S.~Abdelhameed}$^\textrm{\scriptsize 85b}$,
\AtlasOrcid[0000-0002-1002-1652]{K.~Abeling}$^\textrm{\scriptsize 55}$,
\AtlasOrcid[0000-0001-5763-2760]{N.J.~Abicht}$^\textrm{\scriptsize 49}$,
\AtlasOrcid[0000-0002-8496-9294]{S.H.~Abidi}$^\textrm{\scriptsize 30}$,
\AtlasOrcid[0009-0003-6578-220X]{M.~Aboelela}$^\textrm{\scriptsize 45}$,
\AtlasOrcid[0000-0002-9987-2292]{A.~Aboulhorma}$^\textrm{\scriptsize 36e}$,
\AtlasOrcid[0000-0001-5329-6640]{H.~Abramowicz}$^\textrm{\scriptsize 156}$
% ...
\bigskip
\\

% ...
$^{17}$Department for Physics and Technology, University of Bergen, Bergen; Norway.\\
% ...
$^{30}$Physics Department, Brookhaven National Laboratory, Upton NY; United States of America.\\
% ...
$^{36}$$^{(a)}$Facult\'e des Sciences Ain Chock, Universit\'e Hassan II de Casablanca;$^{(b)}$Facult\'{e} des Sciences, Universit\'{e} Ibn-Tofail, K\'{e}nitra;$^{(c)}$Facult\'e des Sciences Semlalia, Universit\'e Cadi Ayyad, LPHEA-Marrakech;$^{(d)}$LPMR, Facult\'e des Sciences, Universit\'e Mohamed Premier, Oujda;$^{(e)}$Facult\'e des sciences, Universit\'e Mohammed V, Rabat;$^{(f)}$Institute of Applied Physics, Mohammed VI Polytechnic University, Ben Guerir; Morocco.\\
% ...
$^{45}$Physics Department, Southern Methodist University, Dallas TX; United States of America.\\
% ...
$^{49}$Fakult\"{a}t Physik, Technische Universit{\"a}t Dortmund, Dortmund; Germany.\\
% ...
$^{55}$II. Physikalisches Institut, Georg-August-Universit\"{a}t G\"ottingen, G\"ottingen; Germany.\\
% ...
$^{85}$$^{(a)}$Khalifa University of Science and Technology, Abu Dhabi;$^{(b)}$New York University Abu Dhabi, Abu Dhabi;$^{(c)}$United Arab Emirates University, Al Ain;$^{(d)}$University of Sharjah, Sharjah; United Arab Emirates.\\
% ...
$^{104}$CPPM, Aix-Marseille Universit\'e, CNRS/IN2P3, Marseille; France.\\
% ...
$^{122}$Homer L. Dodge Department of Physics and Astronomy, University of Oklahoma, Norman OK; United States of America.\\
% ...
$^{156}$Raymond and Beverly Sackler School of Physics and Astronomy, Tel Aviv University, Tel Aviv; Israel.\\
% ...
$^{a}$ Also at Affiliated with an institute formerly covered by a cooperation agreement with CERN.\\
$^{b}$ Also at An-Najah National University, Nablus; Palestine.\\
$^{c}$ Also at Borough of Manhattan Community College, City University of New York, New York NY; United States of America.\\
$^{d}$ Also at Center for Interdisciplinary Research and Innovation (CIRI-AUTH), Thessaloniki; Greece.\\
$^{e}$ Also at Centre of Physics of the Universities of Minho and Porto (CF-UM-UP); Portugal.\\
$^{f}$ Also at CERN, Geneva; Switzerland.\\
% ...

\end{flushleft}


%% file: ANA-HION-2023-03-PAPER.bib
@article{Miller:2007ri,
    author = "Miller, Michael L. and Reygers, Klaus and Sanders, Stephen J. and Steinberg, Peter",
    title = "{Glauber modeling in high energy nuclear collisions}",
    eprint = "nucl-ex/0701025",
    archivePrefix = "arXiv",
    doi = "10.1146/annurev.nucl.57.090506.123020",
    journal = "Ann. Rev. Nucl. Part. Sci.",
    volume = "57",
    pages = "205--243",
    year = "2007"
}

@article{GEANT4:2002zbu,
    author = "Agostinelli, S. and others",
    collaboration = "GEANT4",
    title = "{GEANT4--a simulation toolkit}",
    reportNumber = "SLAC-PUB-9350, FERMILAB-PUB-03-339, CERN-IT-2002-003",
    doi = "10.1016/S0168-9002(03)01368-8",
    journal = "Nucl. Instrum. Meth. A",
    volume = "506",
    pages = "250--303",
    year = "2003"
}

@article{Zapp:2008gi,
    author = "Zapp, Korinna and Ingelman, Gunnar and Rathsman, Johan and Stachel, Johanna and Wiedemann, Urs Achim",
    title = "{A Monte Carlo model for 'jet quenching'}",
    eprint = "0804.3568",
    archivePrefix = "arXiv",
    primaryClass = "hep-ph",
    reportNumber = "CERN-PH-TH-2008-067",
    doi = "10.1140/epjc/s10052-009-0941-2",
    journal = "Eur. Phys. J. C",
    volume = "60",
    pages = "617--632",
    year = "2009"
}

@article{Zapp:2013vla,
    author = "Zapp, Korinna",
    title = "{JEWEL 2.0.0: directions for use}",
    eprint = "1311.0048",
    archivePrefix = "arXiv",
    primaryClass = "hep-ph",
    reportNumber = "CERN-PH-TH-2013-255, MCNET-13-17",
    doi = "10.1140/epjc/s10052-014-2762-1",
    journal = "Eur. Phys. J. C",
    volume = "74",
    number = "2",
    pages = "2762",
    year = "2014"
}

@article{JETSCAPE:2023hqn,
    author = "{JETSCAPE Collaboration}",
    collaboration = "JETSCAPE",
    title = "{Hard jet substructure in a multistage approach}",
    eprint = "2301.02485",
    archivePrefix = "arXiv",
    primaryClass = "hep-ph",
    doi = "10.1103/PhysRevC.110.044907",
    journal = "Phys. Rev. C",
    volume = "110",
    number = "4",
    pages = "044907",
    year = "2024"
}

@article{KunnawalkamElayavalli:2016ttl,
    author = "Kunnawalkam Elayavalli, Raghav and Zapp, Korinna Christine",
    title = "{Simulating V+jet processes in heavy ion collisions with JEWEL}",
    eprint = "1608.03099",
    archivePrefix = "arXiv",
    primaryClass = "hep-ph",
    reportNumber = "CERN-TH-2016-183, MCNET-16-33",
    doi = "10.1140/epjc/s10052-016-4534-6",
    journal = "Eur. Phys. J. C",
    volume = "76",
    number = "12",
    pages = "695",
    year = "2016"
}

@article{Busza:2018rrf,
    author = "Busza, Wit and Rajagopal, Krishna and van der Schee, Wilke",
    title = "{Heavy Ion Collisions: The Big Picture, and the Big Questions}",
    eprint = "1802.04801",
    archivePrefix = "arXiv",
    primaryClass = "hep-ph",
    reportNumber = "MIT-CTP-4892, MIT-CTP/4892",
    doi = "10.1146/annurev-nucl-101917-020852",
    journal = "Ann. Rev. Nucl. Part. Sci.",
    volume = "68",
    pages = "339--376",
    year = "2018"
}

@article{Cunqueiro:2021wls,
    author = "Cunqueiro, Leticia and Sickles, Anne M.",
    title = "{Studying the QGP with Jets at the LHC and RHIC}",
    eprint = "2110.14490",
    archivePrefix = "arXiv",
    primaryClass = "nucl-ex",
    doi = "10.1016/j.ppnp.2022.103940",
    journal = "Prog. Part. Nucl. Phys.",
    volume = "124",
    pages = "103940",
    year = "2022"
}

@article{Casalderrey-Solana:2012evi,
    author = "Casalderrey-Solana, Jorge and Mehtar-Tani, Yacine and Salgado, Carlos A. and Tywoniuk, Konrad",
    title = "{New picture of jet quenching dictated by color coherence}",
    eprint = "1210.7765",
    archivePrefix = "arXiv",
    primaryClass = "hep-ph",
    reportNumber = "CERN-PH-TH-2012-291, LU-TP-12-37",
    doi = "10.1016/j.physletb.2013.07.046",
    journal = "Phys. Lett. B",
    volume = "725",
    pages = "357--360",
    year = "2013"
}

@article{Casalderrey-Solana:2019ubu,
    author = "Casalderrey-Solana, J. and Milhano, G. and Pablos, D. and Rajagopal, K.",
    title = "{Modification of Jet Substructure in Heavy Ion Collisions as a Probe of the Resolution Length of Quark-Gluon Plasma}",
    eprint = "1907.11248",
    archivePrefix = "arXiv",
    primaryClass = "hep-ph",
    doi = "10.1007/JHEP01(2020)044",
    journal = "JHEP",
    volume = "01",
    pages = "044",
    year = "2020"
}

@article{JETSCAPE:2022hcb,
    author = "{JETSCAPE Collaboration}",
    collaboration = "JETSCAPE",
    title = "{Multiscale evolution of charmed particles in a nuclear medium}",
    eprint = "2208.00983",
    archivePrefix = "arXiv",
    primaryClass = "nucl-th",
    doi = "10.1103/PhysRevC.107.054901",
    journal = "Phys. Rev. C",
    volume = "107",
    number = "5",
    pages = "054901",
    year = "2023"
}

@article{Luo:2018pto,
    author = "Luo, Tan and Cao, Shanshan and He, Yayun and Wang, Xin-Nian",
    title = "{Multiple jets and $\gamma$-jet correlation in high-energy heavy-ion collisions}",
    eprint = "1803.06785",
    archivePrefix = "arXiv",
    primaryClass = "hep-ph",
    doi = "10.1016/j.physletb.2018.06.025",
    journal = "Phys. Lett. B",
    volume = "782",
    pages = "707--716",
    year = "2018"
}

@article{Alwall:2014hca,
    author = "Alwall, J. and Frederix, R. and Frixione, S. and Hirschi, V. and Maltoni, F. and Mattelaer, O. and Shao, H. -S. and Stelzer, T. and Torrielli, P. and Zaro, M.",
    title = "{The automated computation of tree-level and next-to-leading order differential cross sections, and their matching to parton shower simulations}",
    eprint = "1405.0301",
    archivePrefix = "arXiv",
    primaryClass = "hep-ph",
    reportNumber = "CERN-PH-TH-2014-064, CP3-14-18, LPN14-066, MCNET-14-09, ZU-TH-14-14",
    doi = "10.1007/JHEP07(2014)079",
    journal = "JHEP",
    volume = "07",
    pages = "079",
    year = "2014"
}

@article{JETSCAPE:2022jer,
    author = "{JETSCAPE Collaboration}",
    collaboration = "JETSCAPE",
    title = "{Inclusive jet and hadron suppression in a multistage approach}",
    eprint = "2204.01163",
    archivePrefix = "arXiv",
    primaryClass = "hep-ph",
    doi = "10.1103/PhysRevC.107.034911",
    journal = "Phys. Rev. C",
    volume = "107",
    number = "3",
    pages = "034911",
    year = "2023"
}

@article{Habich:2014jna,
    author = "Habich, M. and Nagle, J. L. and Romatschke, P.",
    title = "{Particle spectra and HBT radii for simulated central nuclear collisions of C + C, Al + Al, Cu + Cu, Au + Au, and Pb + Pb from $\sqrt{s}=62.4$ - $2760$ GeV}",
    eprint = "1409.0040",
    archivePrefix = "arXiv",
    primaryClass = "nucl-th",
    doi = "10.1140/epjc/s10052-014-3206-7",
    journal = "Eur. Phys. J. C",
    volume = "75",
    number = "1",
    pages = "15",
    year = "2015"
}

@inproceedings{Adye:2011gm,
    author = "Adye, Tim",
    title = "{Unfolding algorithms and tests using RooUnfold}",
    booktitle = "{PHYSTAT 2011}",
    eprint = "1105.1160",
    archivePrefix = "arXiv",
    primaryClass = "physics.data-an",
    doi = "10.5170/CERN-2011-006.313",
    publisher = "CERN",
    address = "Geneva",
    pages = "313--318",
    year = "2011"
}

@article{Sjostrand:2007gs,
    author = "Sjostrand, Torbjorn and Mrenna, Stephen and Skands, Peter",
    title = "{A Brief Introduction to PYTHIA 8.1}",
    eprint = "0710.3820",
    archivePrefix = "arXiv",
    primaryClass = "hep-ph",
    reportNumber = "CERN-LCGAPP-2007-04, LU-TP-07-28, FERMILAB-PUB-07-512-CD-T",
    doi = "10.1016/j.cpc.2008.01.036",
    journal = "Comput. Phys. Commun.",
    volume = "178",
    pages = "852--867",
    year = "2008"
}

@article{Cacciari:2008gp,
    author = "Cacciari, Matteo and Salam, Gavin P. and Soyez, Gregory",
    title = "{The anti-$k_t$ jet clustering algorithm}",
    eprint = "0802.1189",
    archivePrefix = "arXiv",
    primaryClass = "hep-ph",
    reportNumber = "LPTHE-07-03",
    doi = "10.1088/1126-6708/2008/04/063",
    journal = "JHEP",
    volume = "04",
    pages = "063",
    year = "2008"
}

@article{Cacciari:2011ma,
    author = "Cacciari, Matteo and Salam, Gavin P. and Soyez, Gregory",
    title = "{FastJet User Manual}",
    eprint = "1111.6097",
    archivePrefix = "arXiv",
    primaryClass = "hep-ph",
    reportNumber = "CERN-PH-TH-2011-297",
    doi = "10.1140/epjc/s10052-012-1896-2",
    journal = "Eur. Phys. J. C",
    volume = "72",
    pages = "1896",
    year = "2012"
}

@article{DAgostini:1994fjx,
    author = "D'Agostini, G.",
    title = "{A Multidimensional unfolding method based on Bayes' theorem}",
    reportNumber = "DESY-94-099",
    doi = "10.1016/0168-9002(95)00274-X",
    journal = "Nucl. Instrum. Meth. A",
    volume = "362",
    pages = "487--498",
    year = "1995"
}

@article{ALICE:2015mjv,
    author = "{ALICE Collaboration}",
    collaboration = "ALICE",
    title = "{Measurement of jet suppression in central Pb-Pb collisions at $\sqrt{s_\mathrm{NN}}$ = 2.76 TeV}",
    eprint = "1502.01689",
    archivePrefix = "arXiv",
    primaryClass = "nucl-ex",
    reportNumber = "CERN-PH-EP-2015-023",
    doi = "10.1016/j.physletb.2015.04.039",
    journal = "Phys. Lett. B",
    volume = "746",
    pages = "1--14",
    year = "2015"
}

@article{PHENIX:2001hpc,
    author = "{PHENIX Collaboration}",
    collaboration = "PHENIX",
    title = "{Suppression of hadrons with large transverse momentum in central Au+Au collisions at $\sqrt{s_{NN}}$ = 130-GeV}",
    eprint = "nucl-ex/0109003",
    archivePrefix = "arXiv",
    doi = "10.1103/PhysRevLett.88.022301",
    journal = "Phys. Rev. Lett.",
    volume = "88",
    pages = "022301",
    year = "2002"
}

@article{Dulat:2015mca,
    author = "Dulat, Sayipjamal and Hou, Tie-Jiun and Gao, Jun and Guzzi, Marco and Huston, Joey and Nadolsky, Pavel and Pumplin, Jon and Schmidt, Carl and Stump, Daniel and Yuan, C. P.",
    title = "{New parton distribution functions from a global analysis of quantum chromodynamics}",
    eprint = "1506.07443",
    archivePrefix = "arXiv",
    primaryClass = "hep-ph",
    doi = "10.1103/PhysRevD.93.033006",
    journal = "Phys. Rev. D",
    volume = "93",
    number = "3",
    pages = "033006",
    year = "2016"
}

@article{Bierlich:2022pfr,
    author = "Bierlich, Christian and others",
    title = "{A comprehensive guide to the physics and usage of PYTHIA 8.3}",
    eprint = "2203.11601",
    archivePrefix = "arXiv",
    primaryClass = "hep-ph",
    reportNumber = "LU-TP 22-16, MCNET-22-04, FERMILAB-PUB-22-227-SCD",
    doi = "10.21468/SciPostPhysCodeb.8",
    journal = "SciPost Phys. Codeb.",
    volume = "2022",
    pages = "8",
    year = "2022"
}

@article{Buckley:2014ana,
    author = {Buckley, Andy and Ferrando, James and Lloyd, Stephen and Nordstr{\"o}m, Karl and Page, Ben and R{\"u}fenacht, Martin and Sch{\"o}nherr, Marek and Watt, Graeme},
    title = "{LHAPDF6: parton density access in the LHC precision era}",
    eprint = "1412.7420",
    archivePrefix = "arXiv",
    primaryClass = "hep-ph",
    reportNumber = "GLAS-PPE-2014-05, MCNET-14-29, IPPP-14-111, DCPT-14-222",
    doi = "10.1140/epjc/s10052-015-3318-8",
    journal = "Eur. Phys. J. C",
    volume = "75",
    pages = "132",
    year = "2015"
}

@article{ALargeIonColliderExperiment:2021mqf,
    author = {{ALICE Collaboration}},
    collaboration = "A Large Ion Collider Experiment, ALICE",
    title = "{Measurement of the groomed jet radius and momentum splitting fraction in pp and Pb$-$Pb collisions at $\sqrt{s_{NN}} = 5.02$ TeV}",
    eprint = "2107.12984",
    archivePrefix = "arXiv",
    primaryClass = "nucl-ex",
    reportNumber = "CERN-EP-2021-151",
    doi = "10.1103/PhysRevLett.128.102001",
    journal = "Phys. Rev. Lett.",
    volume = "128",
    number = "10",
    pages = "102001",
    year = "2022"
}


%% file: ATLAS.bib
@Article{PERF-2007-01,
    author         = "{ATLAS Collaboration}",
    title          = "{The ATLAS Experiment at the CERN Large Hadron Collider}",
    journal        = "JINST",
    volume         = "3",
    year           = "2008",
    pages          = "S08003",
    doi            = "10.1088/1748-0221/3/08/S08003",
    primaryClass   = "hep-ex",
}

@Article{HION-2011-02,
    author         = "{ATLAS Collaboration}",
    title          = "{Measurement of the jet radius and transverse momentum dependence of inclusive jet suppression in lead--lead collisions at \(\sqrt{s_{\text{NN}}} = 2.76\,\text{TeV}\) with the ATLAS detector}",
    journal        = "Phys. Lett. B",
    volume         = "719",
    year           = "2013",
    pages          = "220",
    doi            = "10.1016/j.physletb.2013.01.024",
    reportNumber   = "CERN-PH-EP-2012-134",
    eprint         = "1208.1967",
    archivePrefix  = "arXiv",
    primaryClass   = "hep-ex",
}

@Article{STDM-2012-16,
    author         = "{ATLAS Collaboration}",
    title          = "{Measurement of the inclusive isolated prompt photons cross section in \(pp\) collisions at \(\sqrt{s} = 7\,\text{TeV}\) with the ATLAS detector using \(4.6\,\text{fb}^{-1}\)}",
    journal        = "Phys. Rev. D",
    volume         = "89",
    year           = "2014",
    pages          = "052004",
    doi            = "10.1103/PhysRevD.89.052004",
    reportNumber   = "CERN-PH-EP-2013-164",
    eprint         = "1311.1440",
    archivePrefix  = "arXiv",
    primaryClass   = "hep-ex",
}

@Article{HION-2013-06,
    author         = "{ATLAS Collaboration}",
    title          = "{Measurements of the Nuclear Modification Factor for Jets in Pb+Pb Collisions at \(\sqrt{s_{\text{NN}}} = 2.76\,\text{TeV}\) with the ATLAS Detector}",
    journal        = "Phys. Rev. Lett.",
    volume         = "114",
    year           = "2015",
    pages          = "072302",
    doi            = "10.1103/PhysRevLett.114.072302",
    reportNumber   = "CERN-PH-EP-2014-172",
    eprint         = "1411.2357",
    archivePrefix  = "arXiv",
    primaryClass   = "hep-ex",
}

@Article{STDM-2014-09,
    author         = "{ATLAS Collaboration}",
    title          = "{Measurement of the inclusive isolated prompt photon cross section in \(pp\) collisions at \(\sqrt{s} = 8\,\text{TeV}\) with the ATLAS detector}",
    journal        = "JHEP",
    volume         = "08",
    year           = "2016",
    pages          = "005",
    doi            = "10.1007/JHEP08(2016)005",
    reportNumber   = "CERN-EP-2016-035",
    eprint         = "1605.03495",
    archivePrefix  = "arXiv",
    primaryClass   = "hep-ex",
}

@Article{STDM-2015-11,
    author         = "{ATLAS Collaboration}",
    title          = "{High-\(E_{\text{T}}\) isolated-photon plus jets production in \(pp\) collisions at \(\sqrt{s} = 8\,\text{TeV}\) with the ATLAS detector}",
    journal        = "Nucl. Phys. B",
    volume         = "918",
    year           = "2017",
    pages          = "257",
    doi            = "10.1016/j.nuclphysb.2017.03.006",
    reportNumber   = "CERN-EP-2016-252",
    eprint         = "1611.06586",
    archivePrefix  = "arXiv",
    primaryClass   = "hep-ex",
}

@Article{HION-2016-07,
    author         = "{ATLAS Collaboration}",
    title          = "{Prompt and non-prompt \(J/\psi\) and \(\psi(2\text{S})\) suppression at high transverse momentum in \(5.02\,\text{TeV}\) Pb+Pb collisions with the ATLAS experiment}",
    journal        = "Eur. Phys. J. C",
    volume         = "78",
    year           = "2018",
    pages          = "762",
    doi            = "10.1140/epjc/s10052-018-6219-9",
    reportNumber   = "CERN-EP-2018-049",
    eprint         = "1805.04077",
    archivePrefix  = "arXiv",
    primaryClass   = "nucl-ex",
}

@Article{STDM-2016-08,
    author         = "{ATLAS Collaboration}",
    title          = "{Measurement of the cross section for inclusive isolated-photon production in \(pp\) collisions at \(\sqrt{s} = 13\,\text{TeV}\) using the ATLAS detector}",
    journal        = "Phys. Lett. B",
    volume         = "770",
    year           = "2017",
    pages          = "473",
    doi            = "10.1016/j.physletb.2017.04.072",
    reportNumber   = "CERN-EP-2016-291",
    eprint         = "1701.06882",
    archivePrefix  = "arXiv",
    primaryClass   = "hep-ex",
}

@Article{TRIG-2016-01,
    author         = "{ATLAS Collaboration}",
    title          = "{Performance of the ATLAS trigger system in 2015}",
    journal        = "Eur. Phys. J. C",
    volume         = "77",
    year           = "2017",
    pages          = "317",
    doi            = "10.1140/epjc/s10052-017-4852-3",
    reportNumber   = "CERN-EP-2016-241",
    eprint         = "1611.09661",
    archivePrefix  = "arXiv",
    primaryClass   = "hep-ex",
}

@Article{HION-2017-08,
    author         = "{ATLAS Collaboration}",
    title          = "{Comparison of Fragmentation Functions for Jets Dominated by Light Quarks and Gluons from \(pp\) and Pb+Pb Collisions in ATLAS}",
    journal        = "Phys. Rev. Lett.",
    volume         = "123",
    year           = "2019",
    pages          = "042001",
    doi            = "10.1103/PhysRevLett.123.042001",
    reportNumber   = "CERN-EP-2019-009",
    eprint         = "1902.10007",
    archivePrefix  = "arXiv",
    primaryClass   = "nucl-ex",
}

@Article{HION-2017-10,
    author         = "{ATLAS Collaboration}",
    title          = "{Measurement of the nuclear modification factor for inclusive jets in Pb+Pb collisions at \(\sqrt{s_{\text{NN}}} = 5.02\,\text{TeV}\) with the ATLAS detector}",
    journal        = "Phys. Lett. B",
    volume         = "790",
    year           = "2019",
    pages          = "108",
    doi            = "10.1016/j.physletb.2018.10.076",
    reportNumber   = "CERN-EP-2018-105",
    eprint         = "1805.05635",
    archivePrefix  = "arXiv",
    primaryClass   = "nucl-ex",
}

@Article{PERF-2017-02,
    author         = "{ATLAS Collaboration}",
    title          = "{Measurement of the photon identification efficiencies with the ATLAS detector using LHC Run~2 data collected in 2015 and 2016}",
    journal        = "Eur. Phys. J. C",
    volume         = "79",
    year           = "2019",
    pages          = "205",
    doi            = "10.1140/epjc/s10052-019-6650-6",
    reportNumber   = "CERN-EP-2018-216",
    eprint         = "1810.05087",
    archivePrefix  = "arXiv",
    primaryClass   = "hep-ex",
}

@Article{PERF-2017-03,
    author         = "{ATLAS Collaboration}",
    title          = "{Electron and photon energy calibration with the ATLAS detector using 2015--2016 LHC proton--proton collision data}",
    journal        = "JINST",
    volume         = "14",
    year           = "2019",
    pages          = "P03017",
    doi            = "10.1088/1748-0221/14/03/P03017",
    reportNumber   = "CERN-EP-2018-296",
    eprint         = "1812.03848",
    archivePrefix  = "arXiv",
    primaryClass   = "hep-ex",
}

@Article{STDM-2017-32,
    author         = "{ATLAS Collaboration}",
    title          = "{Measurement of isolated-photon plus two-jet production in \(pp\) collisions at \(\sqrt{s} = 13\,\text{TeV}\) with the ATLAS detector}",
    journal        = "JHEP",
    volume         = "03",
    year           = "2020",
    pages          = "179",
    doi            = "10.1007/JHEP03(2020)179",
    reportNumber   = "CERN-EP-2019-210",
    eprint         = "1912.09866",
    archivePrefix  = "arXiv",
    primaryClass   = "hep-ex",
}

@Article{EGAM-2018-01,
    author         = "{ATLAS Collaboration}",
    title          = "{Electron and photon performance measurements with the ATLAS detector using the 2015--2017 LHC proton--proton collision data}",
    journal        = "JINST",
    volume         = "14",
    year           = "2019",
    pages          = "P12006",
    doi            = "10.1088/1748-0221/14/12/P12006",
    reportNumber   = "CERN-EP-2019-145",
    eprint         = "1908.00005",
    archivePrefix  = "arXiv",
    primaryClass   = "hep-ex",
}

@Article{HION-2018-06,
    author         = "{ATLAS Collaboration}",
    title          = "{Measurement of photon-jet transverse momentum correlations in \(5.02\,\text{TeV}\) Pb+Pb and \(pp\) collisions with ATLAS}",
    journal        = "Phys. Lett. B",
    volume         = "789",
    year           = "2019",
    pages          = "167",
    doi            = "10.1016/j.physletb.2018.12.023",
    reportNumber   = "CERN-EP-2018-196",
    eprint         = "1809.07280",
    archivePrefix  = "arXiv",
    primaryClass   = "nucl-ex",
}

@Article{TRIG-2018-05,
    author         = "{ATLAS Collaboration}",
    title          = "{Performance of electron and photon triggers in ATLAS during LHC Run~2}",
    journal        = "Eur. Phys. J. C",
    volume         = "80",
    year           = "2020",
    pages          = "47",
    doi            = "10.1140/epjc/s10052-019-7500-2",
    reportNumber   = "CERN-EP-2019-169",
    eprint         = "1909.00761",
    archivePrefix  = "arXiv",
    primaryClass   = "hep-ex",
}

@Article{HION-2019-02,
    author         = "{ATLAS Collaboration}",
    title          = "{Measurements of the suppression and correlations of dijets in Pb+Pb collisions at \(\sqrt{s_{\text{NN}}} = 5.02\,\text{TeV}\)}",
    journal        = "Phys. Rev. C",
    volume         = "107",
    year           = "2023",
    pages          = "054908",
    doi            = "10.1103/PhysRevC.107.054908",
    reportNumber   = "CERN-EP-2022-046",
    eprint         = "2205.00682",
    archivePrefix  = "arXiv",
    primaryClass   = "nucl-ex",
    related        = "HION-2019-02-err",
    relatedstring  = "Erratum:",
}

@Article{HION-2019-11,
    author         = "{ATLAS Collaboration}",
    title          = "{Measurement of azimuthal anisotropy of muons from charm and bottom hadrons in Pb+Pb collisions at \(\sqrt{s_{\text{NN}}} = 5.02\,\text{TeV}\) with the ATLAS detector}",
    journal        = "Phys. Lett. B",
    volume         = "807",
    year           = "2020",
    pages          = "135595",
    doi            = "10.1016/j.physletb.2020.135595",
    reportNumber   = "CERN-EP-2019-274",
    eprint         = "2003.03565",
    archivePrefix  = "arXiv",
    primaryClass   = "nucl-ex",
}

@Article{JETM-2019-01,
    author         = "{ATLAS Collaboration}",
    title          = "{Energy scale and resolution for anti-\(k_t\) jets with radius parameters \(R=0.2\) and \(0.6\) measured in proton--proton collisions at \(\sqrt{s} = 13\,\text{TeV}\) with the ATLAS detector}",
    journal        = "Eur. Phys. J. C",
    volume         = "85",
    year           = "2025",
    pages          = "791",
    doi            = "10.1140/epjc/s10052-025-14226-6",
    reportNumber   = "CERN-EP-2024-339",
    eprint         = "2412.15783",
    archivePrefix  = "arXiv",
    primaryClass   = "hep-ex",
}

@Article{HION-2020-09,
    author         = "{ATLAS Collaboration}",
    title          = "{Measurements of azimuthal anisotropies of jet production in Pb+Pb collisions at \(\sqrt{s_{\text{NN}}} = 5.02\,\text{TeV}\) with the ATLAS detector}",
    journal        = "Phys. Rev. C",
    volume         = "105",
    year           = "2022",
    pages          = "064903",
    doi            = "10.1103/PhysRevC.105.064903",
    reportNumber   = "CERN-EP-2021-218",
    eprint         = "2111.06606",
    archivePrefix  = "arXiv",
    primaryClass   = "nucl-ex",
}

@Article{HION-2021-09,
    author         = "{ATLAS Collaboration}",
    title          = "{Measurement of substructure-dependent jet suppression in Pb+Pb collisions at \(5.02\,\text{TeV}\) with the ATLAS detector}",
    journal        = "Phys. Rev. C",
    volume         = "107",
    year           = "2023",
    pages          = "054909",
    doi            = "10.1103/PhysRevC.107.054909",
    reportNumber   = "CERN-EP-2022-182",
    eprint         = "2211.11470",
    archivePrefix  = "arXiv",
    primaryClass   = "nucl-ex",
}

@Article{HION-2022-14,
    author         = "{ATLAS Collaboration}",
    title          = "{Comparison of inclusive and photon-tagged jet suppression in \(5.02\,\text{TeV}\) Pb+Pb collisions with ATLAS}",
    journal        = "Phys. Lett. B",
    volume         = "846",
    year           = "2023",
    pages          = "138154",
    doi            = "10.1016/j.physletb.2023.138154",
    reportNumber   = "CERN-EP-2023-029",
    eprint         = "2303.10090",
    archivePrefix  = "arXiv",
    primaryClass   = "nucl-ex",
    related        = "HION-2022-14-err",
    relatedstring  = "Erratum:",
}

@Booklet{ATL-SOFT-PUB-2021-001,
    author         = "{ATLAS Collaboration}",
    title          = "{The ATLAS Collaboration Software and Firmware}",
    howpublished   = "{ATL-SOFT-PUB-2021-001}",
    url            = "https://cds.cern.ch/record/2767187",
    year           = "2021",
}

@Booklet{ATL-SOFT-PUB-2026-001,
    author         = "{ATLAS Collaboration}",
    title          = "{ATLAS Computing Acknowledgements}",
    howpublished   = "{ATL-SOFT-PUB-2026-001}",
    url            = "https://cds.cern.ch/record/2952666",
    year           = "2026",
}


%% file: CMS.bib
@Article{CMS-HIN-14-010,
    author         = "{CMS Collaboration}",
    title          = "{Measurement of transverse momentum relative to dijet systems in PbPb and \(pp\) collisions at \(\sqrt{s_{\text{NN}}} = 2.76\,\text{TeV}\)}",
    journal        = "JHEP",
    volume         = "01",
    year           = "2016",
    pages          = "006",
    doi            = "10.1007/JHEP01(2016)006",
    reportNumber   = "CERN-PH-EP-2015-260",
    eprint         = "1509.09029",
    archivePrefix  = "arXiv",
    primaryClass   = "hep-ex",
}

@Article{CMS-HIN-16-002,
    author         = "{CMS Collaboration}",
    title          = "{Study of jet quenching with isolated-photon+jet correlations in PbPb and \(pp\) collisions at \(\sqrt{s_{\text{NN}}} = 5.02\,\text{TeV}\)}",
    journal        = "Phys. Lett. B",
    volume         = "785",
    year           = "2018",
    pages          = "14",
    doi            = "10.1016/j.physletb.2018.07.061",
    reportNumber   = "CERN-EP-2017-285",
    eprint         = "1711.09738",
    archivePrefix  = "arXiv",
    primaryClass   = "hep-ex",
}

@Article{CMS-HIN-16-006,
    author         = "{CMS Collaboration}",
    title          = "{Measurement of the Splitting Function in \(pp\) and Pb--Pb Collisions at \(\sqrt{s_{\text{NN}}} = 5.02\,\text{TeV}\)}",
    journal        = "Phys. Rev. Lett.",
    volume         = "120",
    year           = "2018",
    pages          = "142302",
    doi            = "10.1103/PhysRevLett.120.142302",
    reportNumber   = "CERN-EP-2017-205",
    eprint         = "1708.09429",
    archivePrefix  = "arXiv",
    primaryClass   = "hep-ex",
}

@Article{CMS-HIN-18-014,
    author         = "{CMS Collaboration}",
    title          = "{First measurement of large area jet transverse momentum spectra in heavy-ion collisions}",
    journal        = "JHEP",
    volume         = "05",
    year           = "2021",
    pages          = "284",
    doi            = "10.1007/JHEP05(2021)284",
    reportNumber   = "CERN-EP-2020-226",
    eprint         = "2102.13080",
    archivePrefix  = "arXiv",
    primaryClass   = "hep-ex",
}

@Article{CMS-HIN-23-001,
    author         = "{CMS Collaboration}",
    title          = "{Girth and groomed radius of jets recoiling against isolated photons in lead--lead and proton--proton collisions at \(\sqrt{s_{\text{NN}}} = 5.02\,\text{TeV}\)}",
    journal        = "Phys. Lett. B",
    volume         = "861",
    year           = "2025",
    pages          = "139088",
    doi            = "10.1016/j.physletb.2024.139088",
    reportNumber   = "CERN-EP-2024-073",
    eprint         = "2405.02737",
    archivePrefix  = "arXiv",
    primaryClass   = "nucl-ex",
}


%% file: PubNotes.bib
@Booklet{ATL-PHYS-PUB-2014-021,
    author         = "{ATLAS Collaboration}",
    title          = "{ATLAS Pythia~8 tunes to \(7\,\text{TeV}\) data}",
    howpublished   = "{ATL-PHYS-PUB-2014-021}",
    url            = "https://cds.cern.ch/record/1966419",
    year           = "2014",
}
